\documentclass[prd,twocolumn,showpacs,nofootinbib]{revtex4}

\usepackage{amsfonts}
\usepackage{amsmath}
\usepackage{amssymb}
\usepackage{upgreek}
\usepackage{bm}
\usepackage{dcolumn}
\usepackage{epsfig}
\usepackage{graphicx}
\usepackage{graphics}
\usepackage[latin1]{inputenc}
\usepackage{latexsym}
\usepackage{rotating}
\usepackage{hyperref}
\usepackage{xspace} % Sensible space treatment at end of simple macros
\usepackage[usenames,dvipsnames]{color}

\usepackage{ulem}
\definecolor {darkgreen}{rgb}{0.2,0.7,0.2} 

\usepackage{pdfpages}

\newcommand\be{\begin{equation}}
\newcommand\ba{\begin{eqnarray}}
\newcommand\ee{\end{equation}}
\newcommand\ea{\end{eqnarray}}

\newcommand{\nn}{\nonumber}

\newcommand{\LL}{{\mbox{\tiny L}}}
\newcommand{\V}{{\mbox{\tiny V}}}
\newcommand{\R}{{\mbox{\tiny R}}}
\newcommand{\LiL}{{\mbox{\tiny LL}}}
\newcommand{\TT}{{\mbox{\tiny TT}}}
\newcommand{\sn}{{\mbox{\tiny sn}}}
\newcommand{\se}{{\mbox{\tiny se}}}
\newcommand{\bb}{{\mbox{\tiny b}}}

\newcommand{\PN}{{\mbox{\tiny PN}}}
\newcommand{\BD}{{\mbox{\tiny BD}}}

\newcommand{\GR}{{\mbox{\tiny GR}}}
\newcommand{\MD}{{\mbox{\tiny MD}}}
\newcommand{\SD}{{\mbox{\tiny SD}}}
\newcommand{\ppE}{{\mbox{\tiny ppE}}}

\newcommand{\beq}{\begin{equation}}
\newcommand{\eeq}{\end{equation}}
\newcommand{\bes}{\begin{subequations}}
\newcommand{\ees}{\end{subequations}}
\newcommand{\beqn}{\begin{eqnarray*}}
\newcommand{\eeqn}{\end{eqnarray*}}

\def\nn{\nonumber}

\begin{document}

\title{Model-Independent Test of General Relativity: 
\\ An Extended post-Einsteinian Framework with Complete Polarization Content}
%OTHER OPTIONS:
%Improved post-Einsteinian framework:
%Complete polarization content, many detectors case

\author{Katerina Chatziioannou}
\affiliation{Department of Physics, Montana State University, Bozeman, MT 59718, USA.}
\author{Nicol\'as Yunes}
\affiliation{Department of Physics, Montana State University, Bozeman, MT 59718, USA.}
\author{Neil Cornish}
\affiliation{Department of Physics, Montana State University, Bozeman, MT 59718, USA.}

\date{\today}

%%%%%%%%%%%%%%%%%%%%%%%%%%%%%%%%%%%%%%%%%%%%%%%
\begin{abstract}

We develop a model-independent test of General Relativity that allows for the constraint of the gravitational wave (GW) polarization content with GW detections of binary compact object inspirals. We first consider three modified gravity theories (Brans-Dicke theory, Rosen's theory and Lightman-Lee theory) and calculate the response function of ground-based detectors to gravitational waves in the inspiral phase. This allows us to see how additional polarizations predicted in these theories modify the General Relativistic prediction of the response function. We then consider general power-law modifications to the Hamiltonian and radiation-reaction force and study how these modify the time-domain and Fourier response function when all polarizations are present. From these general arguments and specific modified gravity examples, we infer an improved parameterized post-Einsteinian template family with complete polarization content. This family enhances General Relativity templates through the inclusion of new theory parameters, reducing to the former when these parameters acquire certain values, and recovering modified gravity predictions for other values, including all polarizations. We conclude by discussing detection strategies to constrain these new, polarization theory parameters by constructing certain null channels through the combination of output from multiple detectors.  

\end{abstract}
%%%%%%%%%%%%%%%%%%%%%%%%%%%%%%%%%%%%%%%%%%%%%%%

\maketitle

%%%%%%%%%%%%%%%%%%%%%%%%%%%%%%%%%%%%%%%%%%%%%%%
\section{Introduction}
\label{intro}

Gravitational waves (GWs) will soon be detected by ground-based detectors, such as Advanced LIGO~\cite{Abramovici:1992ah,Abbott:2007kva,ligo} and Advanced Virgo~\cite{Acernese:2007zze,virgo}. These waves will provide invaluable information about the gravitational interaction in the so-called {\emph{strong-field}} regime, where the non-linearity and strong dynamics of the Einstein equations play an important role. Although General Relativity (GR) has passed all Solar System and binary pulsar tests with flying colors, the strong-field regime remains mostly unexplored~\cite{Will:1993ns,Will:2005va}. For example, observations have not yet been able to confirm GR's no-hair theorems or the non-linear part of the Einstein equations in the GW generation.    

One of the primary targets of these terrestrial detectors are waves generated in the late inspiral of compact objects. Detectors that can observe signals as low as $10 \; {\rm{Hz}}$ will be able to follow the binary inspiral from astronomically small separations down to merger. For example, for a binary neutron star inspiral, ground-based detectors should be able to detect inspirals from initial separations of $\sim 600 m_{\rm NS} \sim 2500 \; {\rm{km}}$, where $m_{\rm NS}$ is the total mass of the binary (here and in what follows we use geometric units $G=c=1$). In this late inspiral regime, the non-linearities of the field equations and the strong-field nature of GR are essential, while one can still employ perturbation theory to model the associated GWs until right before the compact objects plunge into each other and merge. 

One of the predictions of GR that one would wish to test is that the GW metric perturbation only possesses two propagating degrees of freedom. In a general theory of gravity, there are up to $6$ degrees of freedom allowed. In GR, however, due to the structure of the field equations, only $2$ of them are physical, with the remaining $4$ being gauge degrees of freedom. Although ground-based detectors are expected to observe GWs at low signal-to-noise ratio, the simultaneous detection of such waves by multiple detectors should allow us to constrain the existence of additional polarization modes (2 scalar and 2 vectorial). The main question, of course, is precisely how to carry out such tests and the degree to which they will allow us to rule out or confirm additional polarization modes. 

Many modified gravity theories exist where all $6$ degrees of freedom are physical and must be accounted for when computing the GW response function. Perhaps the most well-known examples are scalar-tensor theories, where the presence of a scalar field leads to the existence of dipole radiation and a scalar (so-called {\emph{breathing}}) polarization mode, in addition to the two standard transverse-traceless modes of GR~\cite{Brans:1961sx,Will:1993ns,Will:2005va}. Vector-tensor theories usually predict the existence of preferred directions and the excitation of vector modes~\cite{Will:1993ns}, while tensor-vector-scalar theories, such as TeVeS~\cite{Bekenstein:2004ne,Bekenstein:2005nv}, predict the existence of all $6$ polarization modes and Einstein-Aether theories~\cite{Jacobson:2004ts,Eling:2004dk,Jacobson:2008aj} predict the existence of $5$ polarization modes . Moreover, bimetric and stratified theories also predict the existence of all $6$ polarization modes~\cite{Rosen:1971qp,Rosen:1974ua,1973PhRvD...8.3293L,Will:1993ns}. 

Recently, a model-independent framework to test GR was developed: the parameterized post-Einsteinian (ppE) framework~\cite{Yunes:2009ke,Cornish:2011ys}. In this scheme, one enhances the GR GW template family through the addition of new theory parameters $\vec{\theta}$; when these ppE parameters acquire certain values one recovers GR, while for other values one recovers predictions from modified gravity theories.  As a first step, Yunes and Pretorius~\cite{Yunes:2009ke} proposed a simple ppE model that treated the inspiral phase in a post-Newtonian (PN) approximation, where the orbital velocity is much less than the speed of light $v/c \ll1$, and that neglected the direct excitation of additional polarization modes in the GW response function. Such modes were only indirectly accounted for through possible modifications to the orbital binding energy and its rate of change, which were captured through ppE parameters. 

In this paper, we relax this assumption and improve the ppE framework to allow for the direct presence of additional polarization modes in the response function. In order to achieve this, we first look at three modified gravity theories that predict the existence of additional polarization modes, namely Brans-Dicke theory~\cite{Brans:1961sx}, Rosen's bimetric theory~\cite{Rosen:1971qp,Rosen:1974ua} and Lightman-Lee theory~\cite{1973PhRvD...8.3293L}\footnote{We choose these theories as toy models because substantial work has already been done to analyze their properties~\cite{Will:1977wq,Will:1989sk}. Once similar work is done for other theories, the analysis of this paper can be extended to include them.}. For each of these theories, we calculate the GW metric perturbation, we extract from this the polarization modes, construct the GW response function and Fourier transform the latter in the stationary-phase approximation (SPA)~\cite{Bender,Droz:1999qx,Yunes:2009yz}. 

We then use physical arguments to predict the general functional form of the time-domain response function, assuming all polarization states are present. We begin by noting that the existence of additional modes leads to new terms in the time-domain response function that are proportional to the $\ell = (0,1,2)$ harmonics of the orbital phase. When this is Fourier transformed, however, the $\ell = 0$ harmonic does not possess a stationary point, and thus, it is subdominant relative to the $\ell = (1,2)$ modes. When Fourier transforming, we allow for a parametric deformation of the Hamiltonian and the radiation-reaction force through the addition of relative power-law corrections. As Yunes and Pretorius found~\cite{Yunes:2009ke}, power-law modifications to the binding energy and energy flux introduce power-law modifications to the Fourier amplitude and phase. The inclusion of additional polarizations introduces a new term that is proportional to the $\ell=1$ harmonic of the orbital phase and that was not included in the original ppE scheme.

With these general arguments and solutions from modified gravity theories at hand, we generalize the ppE framework to allow for the remaining four polarization modes in the response function. Given a single detector, we find that to parameterize only the $\ell=2$ harmonic one requires $4$ ppE parameters for the most general case, in agreement with Yunes and Pretorius~\cite{Yunes:2009ke}; if one wishes to include all polarization modes, one needs a total of $9$ ppE parameters instead. However, if one restricts attention only to power-law modifications to the Hamiltonian and radiation-reaction force in the generation of GWs, then for single-detectors one needs only $2$ ppE parameters to parameterize the $\ell=2$ harmonic and $4$ ppE parameters to parameterize all polarizations. 
   
Part of our work is similar to a paper by Arun that recently appeared in the literature ~\cite{Arun:2012hf}. In that paper, he first defines the \emph{dipolar mode of the GW} as the $\ell=1$ harmonic component of the wave. Then, assuming a general functional form for this component and a GR quadrupole plus dipole frequency evolution, he calculates the Fourier transform of the response function. He finds that the Fourier response can be parametrized by 2 parameters: one that measures the amplitude of the dipole mode relative to the leading quadrupole mode; and one that measures the relative strength of dipole emission. The reason that Arun manages to parameterize the waveform with a smaller number of parameter is that he neglects corrections to the conservative part of the dynamics and he restricts attention to dipole emission only. Our paper generalizes his results, allowing for generic deformations in the Hamiltonian and radiation-reaction force, while simultaneously accounting for all polarization modes.

Using a single spherical~\cite{springerlink:10.1007/BF02450446} or truncated icosohedral~\cite{PhysRevLett.70.2367} acoustic
detector it is possible to extract and separate~\cite{Bianchi:1996pc} the various polarization modes.
In contrast, a standard two-arm interferometric detector only provides a single projection of the
polarization state, and it takes multiple independent projections from an array of detectors to
fully constrain the polarization content. We construct a ppE model that could be used with multiple
interferometers by restoring the beam-pattern dependencies of the ppE amplitude. This forces us to break-up the overall amplitude into several terms, each with new ppE theory parameters, thus enlarging the parameter space to $18$ theory parameters in the general case and $10$ for power-law corrections. With this at hand, we then discuss a new multiple detector strategy to extract the additional polarization modes through the
construction of {\emph{null streams}}, ie.~combinations that isolate specific polarization modes.
One such null stream has the property that if GR were right, the stream would be free of gravitational
wave energy for any designated sky location. If each detector output is thought of as a vector in signal space, then the construction of null channels reduces to the continuous projection of this signal in directions orthogonal to the $2$ GR polarization modes. In general, we find that at least $3$ detectors are necessary to constrain additional modes, $4$ detectors are necessary to constrain the vector modes and with $6$ detectors it is possible to construct streams that are null in every theory of gravity.

The rest of this paper is organized as follows:
Section~\ref{sec:FT-gen-resp} describes how the gravitational response function is computed in a generic modified gravity theory and summarizes the SPA. 
Section~\ref{sec:MG} discusses Brans-Dicke theory, Rosen's theory and Lightman-Lee's theory, computing the response function and its Fourier transform in the SPA for each of them.
Section~\ref{sec:deconstruct} calculates the complete waveform for generic deviations in the systems binding energy and balance law.
Section~\ref{sec:ppEGen} uses the results from the previous sections to construct a generalized ppE framework that allows for the existence of additional polarizations modes.
Section~\ref{sec:null streams} discusses the construction of null streams for generic waveforms with $6$ polarization modes.
Section~\ref{sec:Conclusions} concludes and discusses possible future research directions. 

We follow here the conventions of Misner, Thorne and Wheeler~\cite{Misner:1973cw}: Greek letters stand for spacetime time indices, while Latin letters in the middle of the alphabet $(i,j,k,\ldots)$ stand for spatial indices only; commas in index lists stand for partial derivatives and semi-colons for covariant derivatives; parentheses and square brackets in index lists stand for symmetrization and anti-symmetrization respectively such as $A_{(\alpha \beta)} = (1/2)(A_{\alpha \beta} + A_{\beta \alpha})$ and $A_{[\alpha \beta]} = (1/2)(A_{\alpha \beta} - A_{\beta \alpha})$; the metric is denoted via $g_{\mu \nu}$ with signature $(-,+,+,+)$; the Einstein summation convention and geometric units with $G = c = 1$ is assumed, unless otherwise specified. 

%%%%%%%%%%%%%%%%%%%%%%%%%%%%%%%%%%%%
\section{Fourier Transform of a Generic Response Function}
\label{sec:FT-gen-resp}
In this section, we construct the response function for a generic modified gravity theory, given the GW metric perturbation. We then briefly explain how the SPA to the Fourier transform of this response can be calculated. We conclude the section by showing how the algorithm works in the standard GR limit. 

%--------------------------------------------------------------------------------
\subsection{Polarizations from the Metric Perturbation}
\label{sec:polarizations}

In this subsection, we mainly follow Will~\cite{Will:1993ns,Will:2005va,Will}. The response function of a detector to a wave with all possible polarizations is
\be
\label{response}
h(t)=F_{+}h^{+}+F_{\times}h^{\times}+F_{\se}h^{\se}+F_{\sn}h^{\sn}+F_{\bb}h^{\bb}+F_{\LL}h^{\LL}\,,
\ee
where $F_{\cdot}$ are angular pattern functions and $h^{\cdot}$ are waveform polarizations. The former are given by~\cite{Will}
\begin{align}\label{antpatterns}
F_{+}&=\frac{1}{2}(1+\cos^2{\theta})\cos{2\psi}\cos{2\phi}-\cos{\theta}\sin{2\psi}\sin{2\phi}\,,
\\
F_{\times}&=\frac{1}{2}(1+\cos^2{\theta})\sin{2\psi}\cos{2\phi}+\cos{\theta}\cos{2\psi}\sin{2\phi}\,,
\\
F_{\sn}&=-\sin{\theta}(\cos{\theta}\cos{2\phi}\cos{\psi}-\sin{2\phi}\sin{\psi})\,,
\\
F_{\se}&=-\sin{\theta}(\cos{\theta}\cos{2\phi}\sin{\psi}+\sin{2\phi}\cos{\psi})\,,
\\
F_{\bb}&=-\frac{1}{2}\cos{2\phi}\sin^2{\theta}\,,
\\
F_{\LL}&=\frac{1}{2}\cos{2\phi}\sin^2{\theta}\,.
\end{align}
The waveform polarizations can be computed from the contraction of certain basis vectors $(e_{ij}^{+},e_{ij}^{\times},{e}^{x}_{i},{e}^{y}_{i})$ (see e.g.~\cite{Kidder:1995zr, Will} noting that in~\cite{Will} $({e}^{x}_{i},{e}^{y}_{i})$, ie.~the basis vectors orthogonal to $\hat{N}^{i}$, the unit vector pointing from the source to the detector, are denoted as $(\hat{\theta}_i,\hat{\phi}_i)$) with the waveform amplitudes $A_{\cdot}$, namely
\bes
\label{modes}
\begin{align}
h^{\bb}&=A_{\bb}\,,
\qquad
h^{\LL}=A_{\LL}\,, 
\\
h^{\sn}&={e}^{x}_{i} A^i_{\V}\,,
\qquad
h^{\se}={e}^{y}_{i} A^i_{\V}\,, 
\\
h^{+}&=e^{+}_{ij}A^{ij}_{\TT}\,,
\qquad
h^{\times}=e^{\times}_{ij}A^{ij}_{\TT}\,.
\end{align}
\ees
In these equations, $A_{\bb}$ is the amplitude of the scalar breathing mode, $A_{\LL}$ is the amplitude of the scalar longitudinal mode, $A^{k}_{V}$ are the amplitudes of the vectorial modes and $A^{ij}_{\TT}$ are the amplitudes of the transverse-traceless modes.

The usual way to find the waveform amplitudes $A_{\cdot}$ is to compute the linearized Riemann tensor evaluated with the trace-reversed metric perturbation~\cite{Misner:1973cw,Will}. A more straightforward way to accomplish the same result, however, is to construct operators that act on the trace-reversed metric perturbation directly and return the waveform amplitudes. In terms of these, the amplitudes are given by
\bes
\label{amplitudes}
\ba
A_{\bb}&=&\frac{1}{2}(\hat{N}_j\hat{N}_k {\bar{h}}^{jk}-{\bar{h}}^{00})\,, 
\\ 
A_{\LL}&=&\hat{N}_j\hat{N}_k {\bar{h}}^{jk}+{\bar{h}}^{00}-2\hat{N}_j {\bar{h}}^{0j}\,,
\\
A^k_{\V}&=&P_{j}^{k}(\hat{N}_i {\bar{h}}^{ij}- {\bar{h}}^{0j})\,, 
\\
A^{ij}_{\TT}&=&P_{m}^{i}P^{j}_{l} {\bar{h}}^{ml}-\frac{1}{2}P^{ij}P_{ml} {\bar{h}}^{ml}\,,
\ea
\ees
where $P_{ij}=\delta_{ij}-\hat{N}_i\hat{N}_j$ is a projection operator orthogonal to $\hat{N}^{i}$, a unit vector pointing from the source to the detector, while $\bar{h}^{\mu \nu}$ is the trace-reversed metric perturbation and
$\delta^{ij}$ is the Kronecker delta. 

One might wonder whether we can reconstruct the full metric perturbation from the GW polarization modes in Eq.~\eqref{amplitudes}. Notice, though, that Eq.~\eqref{amplitudes} contains only 6 degrees of freedom (1 in $A_{\bb}$, 1 in $A_{\LL}$, 2 in $A_{\V}^{k}$ because it is transverse and 2 in $A_{\TT}^{ij}$ because it is transverse and traceless), while the full metric perturbation generically contains 10 degrees of freedom. Thus, for the inversion to be unique one must make a gauge choice, such as a pure traceless (yet not fully transverse) gauge. Doing so, the metric perturbation can be written as
\begin{align}
\bar{h}^{00} &= 0\,,
\nonumber \\
\bar{h}^{0i} &= \frac{\hat{N}^{i}}{D} \left(A_{\bb} - \frac{1}{2} A_{\LL} \right)\,,
\nonumber \\
\bar{h}^{ij} &= \frac{3 A_{\bb}}{D} \left(\hat{N}^{i} \hat{N}^{j} - \frac{1}{3} \delta^{ij} \right) +\frac{ 2 \hat{N}^{(i} A_{\V}^{j)}}{D} + \frac{A_{\ij}^{\TT}}{D}\,.
\end{align}
Of course, such a metric reconstruction is unnecessary for our purposes because the observable is the response function and we can project out the relevant degrees of freedom (those in Eq.~\eqref{amplitudes}) without making any gauge choice.

%------------------------------------------------------------
\subsection{Stationary Phase Approximation}
\label{sec:SPA}

In GW data analysis, one often works with the Fourier transform of the response function, which can be obtained analytically in the SPA~\cite{Bender,Droz:1999qx,Yunes:2009yz}. We here briefly review this method. The goal of the SPA is to compute the generalized Fourier integral 
\be
\label{fourier}
\tilde{h}(f)=\int h(t)e^{2\pi i f t}dt\,,
\ee
assuming that the response function $h(t)$ is composed of a slowly varying amplitude $\mathcal{A}(t)$ and a rapidly varying phase $\ell\Phi(t)$, namely
\be
\label{hoft-eq}
h(t) = {\cal{A}}(t) \left(e^{i \ell \Phi(t)} + e^{-i \ell \Phi(t)}\right)\,,
\ee
with $\ell>0$ the harmonic number and $\Phi(t)$ the orbital phase. 

Equation~\eqref{fourier} can be rewritten using Eq.~\eqref{hoft-eq} as
\be
\label{eq:FFT}
\tilde{h}(f)=\int \mathcal{A}(t) \left[e^{2\pi i f t+i\ell\Phi(t)}+e^{2\pi i f t-i\ell\Phi(t)} \right]dt\,.
\ee
The first term in square brackets does not have a stationary point, ie.~a value of $t$ for which the derivative of the argument of the exponential vanishes, $2 \pi f +\ell d\Phi/dt = 0$. Terms without a stationary point contribute subdominantly to the generalized Fourier integral, and thus, they can be neglected by the Riemann-Lebesgue Lemma~\cite{Bender}. The second term in Eq.~\eqref{eq:FFT} does have a stationary point, which after Taylor expanding occurs when the derivative of the argument in the exponential vanishes, ie.~$F(t_0)=f/\ell$, where $F(t)=\dot{\Phi}/(2 \pi)$ is the orbital angular frequency. 

With this at hand, the SPA of $\tilde{h}(f)$ is~\cite{Yunes:2009yz}
\be
\label{spa general}
\tilde{h}(f)=\frac{\mathcal{A}(t_0)}{\sqrt{\ell\dot{F}(t_0)}}e^{-i\Psi}\,,
\ee
where the phase GW $\Psi$ is given by
\be
\label{spa phase general}
\Psi[F(t_0)]=2\pi\int^{F(t_0)}\left(\ell\frac{F'}{\dot{F}'}-\frac{f}{\dot{F}'}\right)dF'+ \frac{\pi}{4}\,.
\ee
In the rest of this paper, we will use these expressions to find an analytic representation of the Fourier transform of the response function in different modified gravity theories.  

As we will see in Sec.~\ref{sec:MG}, the time-domain response function in a generic modified gravity theory will contain terms proportional to the $\ell^{\rm th}$ harmonic of the orbital phase, as shown in Eq.~\eqref{hoft-eq}. Terms in the Fourier transform of the response function proportional to the $\ell=0$ harmonic are of the form  
\be
\label{h3}
\int_{-\infty}^{\infty} F^{2/3} e^{2\pi i f t}dt \sim  \int^{\infty}_{-\infty} \frac{e^{2\pi i f t}}{t^{n}}dt\,,
\ee
where the power $n$ depends on how the frequency evolves. Such an integral vanishes for $0 < n < 1$ when the limits of integration are $\pm\infty$, because, as the complex exponential oscillates, contributions from subsequent intervals cancel out. Indeed, to first order we keep only the leading quadrupole emission for which $n=1/4$. Therefore, we see that the $\ell=0$ harmonic of the orbital phase in the response function will contribute subdominantly to the SPA of the Fourier response. 

%------------------------------------------------------------------
\subsection{General Relativity Limit}

As an illustrative example, let us apply the above formalism to the trace-reversed metric perturbation in GR. The trace-reversed metric perturbation for a two-body, quasi-circular orbit is given, to leading-order in the PN expansion, by (see e.g.~\cite{Blanchet:2002av})
\be
\label{gr wave}
\bar{h}^{ij}=\frac{2}{D} Q^{ij}\,,
\ee
where the quadrupole moment is
\be
Q^{ij}=2\mu\frac{m}{r}(\hat{v}^i\hat{v}^j-\hat{x}^i\hat{x}^j)\,,
\ee
$\mu$ is the reduced mass, $m$ is the total mass, $D$ is the distance to the source, $r$ is the orbital separation and $(\hat{x}^{i}, \hat{v}^{i})$ are orbital trajectory and orbital velocity unit vectors.

In order to explicitly calculate the waveform amplitudes via Eq.~\eqref{amplitudes} we must first express $(\hat{x}^{i},\hat{v}^{i})$ in the source system $(\hat{i}^{i},\hat{j}^{i},\hat{k}^{i})$
\begin{align}
\hat{x}^{i}&=\cos{\Phi} \; \hat{i}^{i} + \sin{\Phi} \; \hat{j}^{i}\,,
\\
\hat{v}^{i}&=-\sin{\Phi} \; \hat{i}^{i} + \cos{\Phi} \; \hat{j}^{i}\,.
\end{align}
Choosing the coordinate system so that the vector from the source to the observer is on the $y$-$z$ plane the vectors $(\hat{N}^{i},{e}^{x}_{i},{e}^{y}_{i})$ and the polarization tensors $e^{+}_{ij},e^{\times}_{ij}$ are 
\begin{align}
\hat{N}^{i} &=\sin{\iota} \; \hat{j}^{i} + \cos{\iota} \;  \hat{k}^{i}\,,
\\
{e}^{x}_{i} &= -\hat{i}^{i}\,,
\\
{e}^{y}_{i} &= \cos{\iota} \; \hat{j}^{i} - \sin{\iota} \; \hat{k}^{i}\,,
\\
e^{+}_{ij} &= \frac{1}{2}({e}^{y}_{i}{e}^{y}_{j}-{e}^{x}_{i}{e}^{x}_{j}) \,,
\\
e^{\times}_{ij} &= \frac{1}{2}({e}^{x}_{i}{e}^{y}_{j}+{e}^{y}_{i}{e}^{x}_{j}) \,,
\end{align}
where we recall that $\Phi$ is the orbital phase and $\iota$ is the inclination angle. Applying the operators in Eq.~\eqref{amplitudes}, we obtain
\be
A^{ij}_{\rm{TT}} =\widehat{\rm{TT}}[\bar{h}^{ij}]\,,
\ee
where $\widehat{\rm{TT}}[\cdot]$ is the transverse-traceless projection operator, and all other amplitudes are zero. We have here imposed the Lorenz gauge condition $\bar{h}^{\mu\nu}{}_{,\nu}=0$, which can be rewritten as $\bar{h}^{\mu 0}{}_{,0}=\hat{N}_j\bar{h}^{\mu j}{}_{,0}$ and $\bar{h}^{00}{}_{,0}=\hat{N}_j \hat{N}_k \bar{h}^{jk}{}_{,0}$~\cite{Will:1993ns,Will:2005va}.

Once we have the amplitudes, we can compute the waveform polarization modes from Eq.~\eqref{modes} to obtain
\begin{align}
\label{gr modes}
h^{+}&=-\frac{2\mu m}{D r}\cos{2\Phi}(1+\cos^2{\iota})\,,
\\
h^{\times}&=-\frac{4\mu m}{Dr}\sin{2\Phi}\cos{\iota}\,,
\end{align}
and all other modes vanish. From Eq.~\eqref{response}, the response is then simply
\be
\label{waveform gr}
h^{\GR}(t)= A_{\GR}\frac{\mathcal{M}}{D}(2\pi \mathcal{M} F)^{2/3} e^{-i2\Phi} + {\rm{c.c.}}\,,
\ee
where
\be
A_{\GR} \equiv -F_{+}(1+\cos^2{\iota}) - 2 i F_{\times}\cos{\iota}\,,
\ee
$2\Phi$ is the GW phase and $\mathcal{M} = \eta^{3/5} m$ is the chirp mass, with $\eta = m_{1} m_{2}/m^{2}$ the symmetric mass ratio. We have here used Kepler's third law to simplify the final result, neglecting subdominant terms in the PN approximation.

The SPA of the Fourier transform of this response function can be computed following Sec.~\ref{sec:SPA}. Using the balance law to relate the rate of change of binding energy to the GW luminosity, we can calculate the frequency evolution, which to leading-order in the PN expansion is given by
\be
\frac{dF}{dt}=\frac{48}{5\pi\mathcal{M}^2}(2\pi\mathcal{M}F)^{11/3} \left[1 + {\cal{O}}(u^{2})\right]\,.
\ee
With this in hand, we can now compute the well-known ({\emph{restricted}}, ie.~leading-order in the amplitude) Fourier transform of the response function in the SPA, namely
\be
\label{eq:GR-SPA}
\tilde{h}^{\GR}(f)= \left(\frac{5 \pi}{96}\right)^{1/2} A_{\GR} \frac{{\cal{M}}^{2}}{D} (\pi \mathcal{M} f)^{-7/6} \; e^{-i\Psi_{\GR}^{(2)}}\,,
\ee
where we have also introduced for future convenience
\be
\Psi_{\GR}^{(\ell)}=-2\pi f t_{c}+\ell\Phi_{c}+\frac{\pi}{4}-\frac{3 \ell}{256 u_{\ell}^5}
\sum_{n=0}^{7} u_{\ell}^{n/3} \left(c_{n}^{\PN}  + l_{n}^{\PN} \ln{u}\right)\,,
\label{eq:PsiGR}
\ee
although only the $\ell=2$ harmonic enters the GR waveform. Here, $(c_{n}^{\PN},l_{n}^{\PN})$ are known PN coefficients that can be read for example from Eq.~$(3.18)$ in~\cite{Buonanno:2009zt}, and we have defined the reduced $\ell$-harmonic frequency 
\be
\label{eq:ul}
u_{\ell}=\left(\frac{2 \pi \mathcal{M} f}{\ell}\right)^{1/3}\,,
\ee 
such that $u_{2} = (\pi {\cal{M}} f)^{1/3}$, with $f$ the GW frequency. 

Up until now, we have concentrate on the restricted PN approximation, but later on it will be important to determine whether modified gravity corrections to the Fourier amplitude are degenerate with PN amplitude corrections in GR. Amplitude corrections arise because the PN waveform contains an infinite number of higher harmonics, as one can see e.g.~in Eq.~$(238)$ of~\cite{Blanchet:2002av}. Therefore, the Fourier transform of such a waveform leads also to a sum of $\ell$ harmonic terms. The dominant one is the $\ell=2$ mode, which was already described above in Eq.~\eqref{eq:GR-SPA}. The next order terms are the $\ell=1$ and $\ell=3$ harmonics, which scale as~\cite{VanDenBroeck:2006qu}
\begin{align}
\label{high-ell-modes-in-GR}
\tilde{h}_{\GR}^{\ell=1} &= \left(\frac{5\pi}{96}\right)^{1/2} A_{\GR}^{(1)} \frac{{\cal{M}}^{2}}{D} 
\eta^{-1/5} \left( \pi {\cal{M}} f \right)^{-5/6} e^{-i \Psi^{(1)}_{\GR}}\,,
\\
\tilde{h}_{\GR}^{\ell=3} &= \left(\frac{5 \pi}{96}\right)^{1/2} A_{\GR}^{(3)} \frac{{\cal{M}}^{2}}{D} 
\eta^{-1/5} \left( \pi {\cal{M}} f \right)^{-5/6} e^{-i \Psi^{(3)}_{\GR}}\,,
\end{align}
where $A_{\GR}^{(1,3)}$ are amplitude factors that depend on different combinations of the inclination and polarization angles. The key point here is that these terms enter at 1PN order higher in the amplitude relative to the dominant $\ell=2$ mode, as can be established by looking at its $f^{-5/6}$ frequency dependence. 

%%%%%%%%%%%%%%%%%%%%%%%%
\section{Modified Gravity Theories}
\label{sec:MG}

%------------------------------------------
\subsection{Brans-Dicke Theory}
\label{sec:BD-theory}

Brans Dicke theory~\cite{Brans:1961sx} is defined by the gravitational action (in Jordan frame)
\be
S_{\BD}=\frac{1}{16\pi} \int d^4x \sqrt{-g} [\phi R-\frac{\omega_{\BD}}{\phi}\phi^{,\mu}\phi_{,\mu}-\phi^2V]\,,
\ee
where $g$ is the determinant of the metric $g_{\mu \nu}$, $R$ is the Ricci scalar, $\phi$ is a dynamical scalar field, $V$ is a potential for the scalar field and $\omega_{\BD}$ is a coupling constant. Usually, one sets the potential to zero, unless one is considering massive Brans-Dicke theory~\cite{Alsing:2011er}. Such a theory is a subset of scalar-tensor theories, where the coupling constant $\omega(\phi) = \omega_{\BD} = {\rm{const}}$. Variation of this action with respect to the metric and the scalar field leads to the modified field equations of the theory. Linearizing these field equations about a flat background $\eta_{\mu \nu}$, one can obtain evolution equations for $h_{\mu \nu}$. We refer the interested reader to~\cite{Will:1989sk,Yunes:2011aa} for further details.

In Brans-Dicke theory, it is convenient to define the trace-reversed metric perturbation in terms of two other fields, a covariantly conserved tensor $\theta^{\mu \nu}$ and the scalar field $\phi$, to obtain
\be
\bar{h}^{\mu \nu}=\theta^{\mu \nu}+\frac{\phi}{\phi_0}\eta^{\mu \nu}\,,
\ee
where $\phi_{0}$ is the asymptotic value of the scalar field at spatial infinity. The linearized field equations prescribe the evolution of both $\phi$ and $\theta^{\mu \nu}$, whose solution in the PN approximation is 
\begin{align}
\theta^{ij}&= \frac{4\mu}{D}\left(1-\frac{1}{2}\xi\right)\frac{Gm}{r}(\hat{v}^i\hat{v}^j-\hat{x}^i\hat{x}^j)\,,
\\
\frac{\phi}{\phi_0}&=-\frac{4\mu}{D}\bar{S}\,,
\end{align}
where we have defined 
\begin{align}
\label{S-def}
\bar{S} &=-\frac{1}{4}\xi \left\{\frac{\Gamma Gm}{r}[(\hat{N}\cdot\hat{v})^2-(\hat{N}\cdot\hat{x})^2]\right.\nn \\
&\left.-(G\Gamma +2\Lambda)\frac{m}{r}-2 S\left(\frac{Gm}{r}\right) ^{1/2} \left(\hat{N}\cdot\hat{v}\right)\right\}\,.
\end{align}
In these equations, we have also defined $\xi=(2+\omega_{\BD})^{-1} \sim \omega_{\BD}^{-1}$ for $\omega_{\BD} \gg 1$, $G=1-\xi(s_1+s_2-2s_1s_2)$, $S=s_1-s_2$, where $s_{A}$ is the \emph{sensitivity} of the $A$th object, as defined in Brans Dicke theory, $\Gamma=1-2(m_1s_2+m_2s_1)/m$ and $\Lambda=1-s_1-s_2 + {\cal{O}}(\xi)$. Clearly, Brans-Dicke theory reduces to GR in the $\omega_{\BD} \to \infty$ (or $\xi \to 0$) limit.

With this at hand, we can now compute the polarization modes as in Sec.~\ref{sec:polarizations}. Using the Lorenz gauge condition from~\cite{Will:1977wq} $\theta^{\mu \nu}{}_{,\nu}=0$\footnote{The evolution equation for $\theta$ is $\square \theta^{\mu \nu} = -16 \pi  \tau^{\mu \nu}$, where $\tau^{\mu \nu}$ is a complicated extension of the Landau-Lifshitz pseudo-tensor. One can easily verify that this differential equation preserves the Lorenz gauge condition.}, the waveform amplitudes are
\be
\label{bdamplitudes}
A_{\bb}=\frac{\phi}{\phi_0}\,, \qquad
A^{ij}_{\TT}=\widehat{\rm{TT}}[\theta^{ij}]\,,
\ee
where the longitudinal and vectorial modes vanish. The polarization modes are then
\ba
\label{bdmodes}
h^{\bb}&=&\frac{-4 \mu \bar{S}}{D}\,,
\\
h^{+}&=&-\left(1-\frac{1}{2}\xi\right) \frac{2 G \mu m}{Dr}\cos{2\Phi}(1+\cos^2{\iota})\,,
\\
h^{\times}&=&-\left(1-\frac{1}{2}\xi\right)\frac{4G \mu  m}{Dr}\sin{2\Phi}\cos{\iota}\,,
\ea
where again the longitudinal and vectorial modes vanish.

Putting all pieces together and simplifying expressions through the leading PN order expression for Kepler's third law
\be\label{keplerbd}
2 \pi F=\left(\frac{G m}{r^3}\right)^{1/2}\,,
\ee
we find 
\ba
\label{waveformbd}
h^{\BD}(t)&=&A_{\BD}\frac{\mathcal{M}}{D}(2\pi \mathcal{M} F)^{2/3} e^{-i 2 \Phi}\nn \\
&+&B_{\BD} \eta^{1/5} \frac{\mathcal{M}}{D}(2\pi \mathcal{M} F)^{1/3}e^{-i \Phi}\nn \\
&+&C_{\BD}\frac{\mathcal{M}}{D}(2\pi \mathcal{M} F)^{2/3}+\textrm{c.c.}\,,
\ea
where ${\rm{c.c.~}}$ stands for the complex conjugate and where we have kept terms only linear in $\xi$. In these equations, we have also defined
\begin{align}
A_{\BD}&=-F_{+}(1+\cos^2{\iota}) - 2iF_{\times}\cos{\iota}
\nn \\
&+ \xi \left[k_{\BD} F_{+}(1+\cos^2{\iota}) +
2 i \; k_{\BD} F_{\times}\cos{\iota}+F_{\bb}\frac{\Gamma}{2}\sin^2{\iota} \right]\,,
\nn \\
&\equiv A_{\GR}+\xi A_{\BD,1}\,,
\\
B_{\BD}&= \xi \left(- F_{\bb}\emph{S} \sin{\iota} \right)=\xi B_{\BD,1}\,,
\\
C_{\BD}&=\xi \left[- \frac{F_{\bb}}{2} (\Gamma+2\Lambda)\right]\,,
\end{align}
and $k_{\BD}= (1/2) + (2/3) (s_1+s_2-2s_1s_2)$.

The Fourier integral of Eq.~\eqref{waveformbd} can be easily calculated with the SPA, but this requires use of the orbital frequency evolution. The rate of change of the binding energy is determined by both dipole and quadrupole radiation, and to first order in $\xi$ and to leading-order in the PN approximation, it is calculated in~\cite{Will:1977wq}
\be
\frac{dE}{dt}=-\frac{8}{15}\frac{\mu^2 m^2}{r^4}\left[12 G^2 \left(1-\frac{1}{2}\xi+\frac{1}{12}\xi \Gamma^2\right)v^2+\frac{5}{4} G^2 \xi S^2\right]\,.
\ee
When this is combined with the binding energy $E =-G m \mu/r$  and Eq.~($\ref{keplerbd}$), one obtains the orbital frequency evolution
\begin{align}
\label{f evolution}
\frac{dF}{dt}&=\frac{48}{5\pi\mathcal{M}^2}(2\pi\mathcal{M}F)^{11/3}
\nn \\
&+ \frac{S^2 \eta^{2/5}}{\pi {\cal{M}}^{2}} \; \xi \; \left(2 \pi {\cal{M}} F\right)^3 
\nn \\
&+ \frac{48}{5\pi\mathcal{M}^2}  \; \xi \;  \left(2\pi\mathcal{M}F\right)^{11/3}\left(\frac{1}{12}\Gamma^2-k_{\BD}\right)\,,
\end{align}
where recall that $G \neq 1$ as given below Eq.~\eqref{S-def}. 

The total Fourier-transformed waveform in Brans-Dicke theory is then simply 
\be
\label{eq:full-BD}
\tilde{h}_{\BD}(f) = \tilde{h}_{\BD}^{(1)}(f) + \tilde{h}_{\BD}^{(2)}(f)\,,
\ee
where the Fourier transform of the first term in Eq.~(\ref{waveformbd}) is
\begin{align}
\label{h1} 
\tilde{h}_{\BD}^{(2)} \hspace{-0.1cm} &= \hspace{-0.1cm} \sqrt{\frac{5 \pi}{96}} \frac{{\cal{M}}^{2}}{D} A_{\BD}   \left[ 1 - \xi \left(\frac{\Gamma^2}{24}-\frac{k_{\BD}}{2}\right) \right] 
\hspace{-0.1cm} \left(\pi {\cal{M}} f\right)^{-7/6} \hspace{-0.1cm} e^{-i\Psi_{\BD}^{(2)}}
\nn \\
&-\xi \left(\frac{5}{96}\right)^{3/2} \hspace{-0.2cm} \pi^{1/2} A_{\BD} S^2\frac{\mathcal{M}^{2}}{D}  \eta^{2/5} 
\left(\pi {\cal{M}} f\right)^{-11/6} e^{-i\Psi_{\BD}^{(2)}}\,,
\end{align}
and that of the second term is
\be\label{h2}
\tilde{h}_{\BD}^{(1)}(f)=\xi B_{\BD,1} \left(\frac{5 \pi}{384}\right)^{1/2} \hspace{-0.2cm}  \frac{\mathcal{M}^{2}}{D} \eta^{1/5} 
\left(\pi {\cal{M}} f\right)^{-3/2} e^{-i\Psi_{\BD}^{(1)}}\,.
\ee
We have here assumed that the second and third terms in Eq.~\eqref{f evolution} are much smaller than the first term, ie.~that Brans-Dicke theories introduces a small deformation away from GR. We have also here defined the $\ell^{\rm th}$-harmonic Brans-Dicke Fourier phase
\begin{align}
\Psi_{\BD}^{(\ell)} &= \Psi_{\GR}^{(\ell)} + \delta\Psi^{(\ell)}_{\BD}\,,
\qquad {\rm{where}}
\nn \\
\delta\Psi_{\BD}^{(\ell)}&=
+\frac{5 \ell }{7168}\xi S^2 \eta^{2/5} u_{\ell}^{-7}\,,
\label{psi1}
\end{align}
with $\Psi_{\GR}^{(\ell)}$ given in Eq.~\eqref{eq:PsiGR} (see also Eq.~\eqref{bdwhole}) and $u_{\ell}$ given by Eq.~\eqref{eq:ul}. Notice that the second term in Eq.~\eqref{h1} is of $-1$PN order relative to the first one, a typical signature of dipole radiation. Such an amplitude correction is usually neglected, because GW interferometers are much more sensitive to the phase evolution.  Of course, not all corrections to the amplitude and the phase will be measurable, and some of them can be degenerate with other system parameters, such as the luminosity distance or the chirp mass. Such issues will be discussed further in Sec.~\ref{sec:ppEGen}.

%------------------------------------------
\subsection{Rosen's Theory}
\label{sec:Rosen}

Rosen's is an example of a bimetric theory~\cite{Rosen:1971qp,Rosen:1974ua}: a theory with a dynamical tensor gravitational field and a flat, non-dynamical metric, or prior geometry. Rosen's theory is defined by the gravitational action~\cite{Rosen:1971qp,Rosen:1974ua,Will:1993ns}
\be
S_{\R} = \frac{1}{32 \pi G} \int d^{4}x \sqrt{-\eta} \eta^{\mu \nu} g^{\alpha \beta} g^{\gamma \delta} \bar{\nabla}_{\mu} g_{\alpha [\gamma} \bar{\nabla}_{|\nu|} g_{\beta] \delta}\,,
\ee
where $\eta$ is the determinant of the flat, non-dynamical metric $\eta_{\mu \nu}$ and the $\bar{\nabla}_{\mu}$ operator stands for a covariant derivative with respect to $\eta_{\mu \nu}$. Although the field equations in this theory are quite different from Einstein's, it has a standard parametrized post-Newtonian (ppN) limit, with only the $\alpha_{2}$ ppN parameter different from its GR value. This parameter, however, has been constrained to be less than $4 \times 10^{-7}$ through observations of solar alignment with the ecliptic plane~\cite{Will:1993ns,Will:2005va}. 

Rosen's theory is of class $II_6$ in the $E(2)$ classification~\cite{Will:1993ns,Will:2005va}, and thus, not all six polarization modes are observer-independent. In theories of this class, all observers agree on the magnitude of the longitudinal mode, but they disagree on the presence or absence of all other modes. This, however, does not mean that the other modes are not real or that they do not carry energy. It only implies that a spin-decomposition of the GWs is not invariant. Such frame dependence is irrelevant for our purposes, since the detector will be in a given frame, and thus, it will measure a certain number of polarization modes. See~\cite{Will:1993ns,Will:2005va} for more details on this theory.

The lack of definite helicity in Rosen's theory is connected to the lack of positive definiteness in the sign of the emitted energy~\cite{1977ApJ...214..826W}. That is, for certain systems, Rosen's theory predicts dipolar radiation that pumps energy into the system, leading to a total energy flux that is positive, instead of negative as in GR. In turn, this for example leads to an increase in the orbital period of binary pulsars with time~\cite{1977ApJ...214..826W,Will:1977zz,1981GReGr..13....1W}. Since binary pulsar observations are consistent with GR, Rosen's theory is today less appealing than in the 1970s. We here study it, not because we think of it as a particularly good candidate to replace Einstein's theory, but as a toy model to determine how the GW response function is modified in theories that allow for the existence of all gravitational polarization modes~\cite{Will:1993ns,Will:2005va}.

The variation of the action and the linearization of the resulting field equations give the evolution of the trace-reversed metric perturbation~\cite{Will:1977wq,Will:1977zz}, which can be solved to find
\begin{align}
\label{rosenwave}
\bar{h}^{00}&=\frac{4\mu}{D}\left\{\frac{m}{r}[(\hat{N}\cdot\hat{v})^2-1-(\hat{N}\cdot\hat{x})^2]+\left(\frac{m}{r}\right)^{1/2}\hspace{-0.2cm}\mathcal{G} \left(\hat{N}\cdot\hat{v}\right)\right\}\,,
\\
\bar{h}^{0j}&=\frac{4\mu}{D}\left\{v^{j}\frac{m}{r}(\hat{N}\cdot\hat{v})-x^{j}\frac{m}{r} \left(\hat{N}\cdot\hat{x}\right)+\frac{2}{3}\left(\frac{m}{r}\right)^{1/2}\mathcal{G}v^{j}\right\}\,,
\\
\bar{h}^{ij}&=\frac{4\mu}{D}\left\{\frac{m}{r}v^{i}v^{j}-\frac{1}{3}\left(\frac{m}{r}\right)^{1/2}\mathcal{G}(\hat{N}\cdot\hat{v})\delta^{ij}\right\}\,,
\end{align}
where $\mathcal{G}$ is the difference in the self-gravitational binding energy per unit mass of the binary components: $\mathcal{G}=s_1/m_1-s_2/m_2$. Due to the specific characteristics of the theory, one cannot find a particular gauge to simplify the above equations. 

With this in hand, we can now compute the polarizations modes and the response function. Following the steps laid out in Sec.~\ref{sec:polarizations}, we find
\ba
\label{rosensmodes}
h^{\bb}&=&\frac{2\mu}{D}\left[\frac{m}{r}\sin^2{\iota}\sin^2{\Phi}+\frac{m}{r}-\frac{4}{3}\left(\frac{m}{r}\right)^{1/2}\mathcal{G}\sin{\iota}\cos{\Phi}\right]\,,
\nn \\
h^{\LL}&=&\frac{4\mu}{D}\left[\frac{m}{r}\sin^2{\iota}\sin^2{\Phi}-\frac{m}{r}-\frac{2}{3}\left(\frac{m}{r}\right)^{1/2}\mathcal{G}\sin{\iota}\cos{\Phi}\right]\,,
\nn \\
h^{\sn}&=&\frac{4\mu}{D}\left[-\frac{m}{r}\cos{\Phi}\sin{\Phi}\sin{\iota}-\frac{2}{3}\left(\frac{m}{r}\right)^{1/2}\mathcal{G}\sin{\Phi}\right]\,,
\nn \\
h^{\se}&=&\frac{4\mu}{D}\left[\frac{m}{r}\sin^2{\Phi}\sin{\iota}\cos{\iota}-\frac{2}{3}\left(\frac{m}{r}\right)^{1/2}\mathcal{G}\cos{\Phi}\sin{\iota}\right]\,,
\nn \\
h^{+}&=&\frac{2\mu m}{Dr}(\sin^2{\Phi}-\cos^2{\iota}\cos^2{\Phi})\,,
\nn \\
h^{\times}&=&-\frac{2\mu m}{Dr}\sin{2\Phi}\cos{\iota}\,.
\ea
Using the modified Kepler's law
\be\label{keplerr}
2 \pi F=\left(k_{\R}\frac{m}{r^3}\right)^{1/2}\,,
\ee 
where $k_{\R} = 1 - 4 s_{1} s_{2}/3$,  the time-domain response function is 
\ba\label{rosenresponse}
h^{\R}(t)&=&A_{\R}\frac{\mathcal{M}}{D} (2\pi \mathcal{M} F)^{2/3} e^{-2i \Phi}
\nn \\
&+&B_{\R}\frac{{\cal{M}}}{D} \eta^{1/5} (2\pi \mathcal{M} F)^{1/3} e^{-i\Phi}
\nn \\
&+&C_{\R}\frac{\mathcal{M}}{D} (2\pi \mathcal{M} F)^{2/3}+\textrm{c.c.}\,,
\ea
where we recall that ${\rm{c.c.}}$ stands for complex conjugate and where $(A_{\R},B_{\R},C_{\R})$ are functions of the angles:
\begin{align}
A_{\R}&=\left(-F_{+}\frac{1+\cos^2{\iota}}{2}-F_{\times}i\cos{\iota}-F_{\bb}\frac{\sin^2{\iota}}{2}\right.\nn \\
&\left.-F_{\LL}\sin^2{\iota}-F_{\sn}i\sin{\iota}-F_{\se}\frac{\sin{2\iota}}{2}\right)k_{\R}^{-1/3}\,,
\\
B_{\R}&=\left(-F_{\bb}\frac{4}{3}\mathcal{G}\sin{\iota}-F_{\LL}\frac{4}{3}\mathcal{G}\sin{\iota}\right. 
\nn\\
&\left.-F_{\sn}\frac{4}{3}i\mathcal{G}-F_{\se}\frac{4}{3}\mathcal{G}\cos{\iota}\right)k_{\R}^{-1/6}\,,
\\
C_{\R}&=\left[F_{+}\frac{\sin^2{\iota}}{2}+F_{\bb}(1+\frac{\sin^2{\iota}}{2})+F_{\se}\frac{\sin{2\iota}}{2}\right.\nn \\
&\left.-F_{\LL}(1+\cos^2{\iota})\right]k_{\R}^{-1/3}\,.
\end{align}
Notice that in the ${\cal{G}} \to 0$ limit, one does not recover GR, as described for example in~\cite{Will:1977wq,Will:1977zz}.\

Before we can calculate the Fourier transform of the response function in the SPA for Rosen's theory, we must first calculate the orbital frequency evolution. The energy evolution of the binary orbit due to GW emission is (to leading-order in the PN approximation and in ${\cal{G}}$) given in ~\cite{Will:1977wq}
\be
\frac{dE}{dt}=\frac{84}{15}\frac{\mu^2 m^2}{r^4}v^2+\frac{20}{9}\frac{\mu^2 m^2 \mathcal{G}^2}{r^4}\,,
\ee
and this leads to the orbital frequency evolution 
\be
\label{rosenfrequency}
\frac{dF}{dt}=-\frac{42 k_{\R}^{-5/6}}{5\pi \mathcal{M}^2}(2\pi\mathcal{M}F)^{11/3}-\frac{10 k_{\R}^{-9/6}}{3\pi {\cal{M}}^{2}} \mathcal{G}^2 \eta^{2/5} (2 \pi {\cal{M}} F)^3\,.
\ee
where the binding energy is not modified.

We can now calculate the Fourier transform in the SPA:
\be
\tilde{h}_{\R}(f) = \tilde{h}_{\R}^{(1)}(f) + \tilde{h}_{\R}^{(2)}(f)\,,
\label{eq:full-h-Rosen}
\ee
where the transform of the first term in Eq.~\eqref{rosenresponse} is 
\begin{align}
\label{eq:Rosen-htilde1}
\tilde{h}_{\R}^{(2)}(f)=A_{\R} k_{\R}^{-5/12}\; i\sqrt{\frac{5 \pi}{84}} \frac{\mathcal{M}^2}{D}(\pi\mathcal{M}f)^{-7/6}e^{-i\Psi_{\R}^{(2)}}\,,
\end{align}
and that of the second term is
\begin{align}
\label{eq:Rosen-htilde2}
\tilde{h}_{\R}^{(1)}(f)=B_{\R}k_{\R}^{-5/12}\; i\sqrt{\frac{5 \pi}{336}} \eta^{1/5} \frac{\mathcal{M}^{2}}{D}(\pi\mathcal{M}f)^{-3/2}e^{-i\Psi_{\R}^{(1)}}\,.
\end{align}
We have here assumed that $\mathcal{G}\ll 1$ and kept terms only to leading-order both in ${\cal{G}}$ and in the PN approximation. We have also here defined the $\ell$-harmonic Rosen Fourier phase 
\begin{eqnarray}
\label{eq:Rosen-Psi1}
&&\Psi_{\R}^{(\ell)}=\frac{\pi}{4}+\ell \Phi_c-2\pi f t_c+\frac{3 \ell }{224u_{\ell}^5}k_{\R}^{-5/6}\nonumber\\
&& \quad +\frac{25 \ell}{8232}\frac{ k_{\R}^{-2/3}\mathcal{G}^2\eta^{2/5}}{u_{\ell}^7}\,,
\end{eqnarray}
where $u_{\ell}$ is given in Eq.~\eqref{eq:ul}. As before, some modifications are degenerate with system parameters, as we will see in Sec.~\ref{sec:ppEGen}.

%------------------------------------------
\subsection{Lightman-Lee Theory}
\label{sec:LLTheory}

Lightman-Lee theory~\cite{1973PhRvD...8.3293L} is a bimetric theory of gravity, similar to Rosen's. This theory is controlled by the metric $g_{\mu \nu}$, a dynamical gravitational tensor $B_{\mu \nu}$ that is connected to $g_{\mu \nu}$ and a flat, non-dynamical background metric~\cite{Will:1993ns}. The theory is defined by the gravitational action
\be
S_{\LiL} = - \frac{1}{16 \pi} \int d^{4}x \sqrt{-\bar{\eta}} \left(\frac{1}{4} B^{\mu \nu|\alpha} B_{\mu \nu|\alpha} - \frac{5}{64} B_{,\alpha} B^{,\alpha} \right)\,,
\ee
where $\bar{\eta}$ is the trace of the background metric $\bar{\eta}_{\mu \nu}$ (not to be confused with the symmetric mass ratio $\eta$ introduced earlier). The spacetime metric is connected to the tensor $B_{\mu \nu}$ via 
\begin{align}
g_{\mu \nu} &= \left(1 - \frac{1}{16} B\right)^{2} \Delta_{\mu}{}^{\alpha} \Delta_{\alpha \nu}\,,
\\
\delta^{\mu}{}_{\nu} &= \Delta^{\alpha}{}_{\nu} \left(\delta_{\alpha}{}^{\mu} - \frac{1}{2} h_{\alpha}{}^{\mu}\right)\,,
\end{align}
which can be expanded for weak gravitational fields as $g_{\mu \nu} = \eta_{\mu \nu} + h_{\mu \nu}$ with $h_{\mu \nu} = B_{\mu \nu} - B \eta_{\mu \nu}/8$. We refer the reader to~\cite{Will:1993ns,Will:2005va} for more details on this theory.

This theory is part of a wider class (sometimes called BSLL) of theories that are semi-conservative. In the particular case of Lightman-Lee theory, there are no preferred frame-effects and the ppN parameters reduce identically to those of GR~\cite{Will:1993ns}. This theory, however, is also of class $II_{6}$ in the $E(2)$ classification, just like Rosen's theory. As before, this implies the lack of definite helicity polarization states and the prediction of the existence of all polarization modes~\cite{1977ApJ...214..826W}.  As in Rosen's theory, lack of helicity seems to lead to a lack of positive definiteness in the energy flux, which is in contradiction with binary pulsars~\cite{1977ApJ...214..826W,1981GReGr..13....1W}. Nonetheless, as in the case of Rosen's theory, we here take Lightman-Lee's as a toy model that allows us to see how the response function is modified in theories that predict all possible GW polarizations.

The trace reversed metric perturbation is given in terms of $B_{\mu \nu}$ by $\bar{h}_{\mu \nu}=B_{\mu \nu}-\frac{3}{8}B\eta_{\mu \nu}$. Using $B$ as given in~\cite{Will:1977wq}, we find that 
\begin{align}
\label{llwave}
\bar{h}^{00}&=\frac{\mu}{D}\left[\frac{m}{r}[2v^2+2(\hat{N}\cdot\hat{v})^2-10-2(\hat{N}\cdot\hat{x})^2]
\right. 
\nn \\
&+ \left. 10\left(\frac{m}{r}\right)^{1/2}\mathcal{G}\hat{N}\cdot\hat{v}\right]\,,\\
\bar{h}^{0j}&=\frac{4\mu}{D}\left[\frac{m}{r}v^{j}(\hat{N}\cdot\hat{v})-x^{j}\frac{m}{r}\hat{N}\cdot\hat{x}+\frac{5}{3}\left(\frac{m}{r}\right)^{1/2}\mathcal{G}v^{j}\right]\,,\\
\bar{h}^{ij}&=\frac{\mu}{D}\left(\{4v^{i}v^{j}-[2v^2-2(\hat{N}\cdot\hat{v})^2+2+2(\hat{N}\cdot\hat{x})^2]\delta^{ij}\}\right.\nn \\
&\times \left. \frac{m}{r} -\frac{10}{3}\left(\frac{m}{r}\right)^{1/2}\mathcal{G}(\hat{N}\cdot\hat{v})\delta^{ij}\right)\,.
\end{align}

As in the case of Rosen's theory, one cannot choose a particular gauge to simplify expressions. We can then extract the polarization modes as explained in Sec.~\ref{sec:polarizations} to obtain
\begin{align}
\label{llmodes}
h^{\bb}&=\frac{\mu}{2D}\left[4\frac{m}{r}\sin^2{\iota}\cos^2{\Phi}-4\frac{m}{r}\sin^2{\iota}\sin^2{\Phi}+4\frac{m}{r}\right.\nn\\
&\left.-\frac{50}{3}\left(\frac{m}{r}\right)^{1/2}\mathcal{G}\sin{\iota}\cos{\Phi}\right]\,,\nn \\
h^{\LL}&=\frac{\mu}{D}\left[4\frac{m}{r}\sin^2{\iota}\sin^2{\Phi}-12\frac{m}{r}
\right.
\nn \\
&\left. -\frac{20}{3}\left(\frac{m}{r}\right)^{1/2}\mathcal{G}\sin{\iota}\cos{\Phi}\right]\,,\nn\\
h^{\sn}&=\frac{\mu}{D}\left[-4\frac{m}{r}\cos{\Phi}\sin{\Phi}\sin{\iota}+\frac{20}{3}\left(\frac{m}{r}\right)^{1/2}\mathcal{G}\sin{\Phi}\right]\,,\nn\\
h^{\se}&=\frac{\mu}{D}\left[4\frac{m}{r}\sin^2{\Phi}\sin{\iota}\cos{\iota}+\frac{20}{3}\left(\frac{m}{r}\right)^{1/2}\mathcal{G}\cos{\Phi}\sin{\iota}\right]\,,\nn\\
h^{+}&=\frac{2\mu m}{Dr}(\sin^2{\Phi}-\cos^2{\iota}\cos^2{\Phi})\,,\nn\\
h^{\times}&=-\frac{2\mu m}{Dr}\sin{2\Phi}\cos{\iota}\,.\nn
\end{align}
With these modes, we can construct the response function to find
\ba
\label{llresponse}
h^{\LiL}(t)&=&A_{\mbox{\tiny LL}}\frac{\mathcal{M}(2\pi \mathcal{M} F)^{2/3}}{D}e^{-2i \Phi}\nn \\
&+&B_{\LiL}\frac{(2\pi \mathcal{M} F)^{1/3}}{D}\mu^{1/2}\mathcal{M}^{1/2}e^{-i\Phi}\nn \\
&+&C_{\LiL}\frac{\mathcal{M}(2\pi \mathcal{M} F)^{2/3}}{D}+ \textrm{c.c.}\,,
\ea
where 
\begin{align}
A_{\LiL}&=-F_{+}\frac{1+\cos^2{\iota}}{2}-F_{\times}i\cos{\iota}+\frac{3}{2}F_{\bb}\sin^2{\iota}\nn \\
&-F_{\LL}\sin^2{\iota}-F_{\sn}i\sin{\iota}-F_{\se}\frac{\sin{2\iota}}{2}\,, \\
B_{\LiL}&=-F_{\bb}\frac{25}{6}\mathcal{G}\sin{\iota}-F_{\LL}\frac{10}{3}\mathcal{G}\sin{\iota}\nn \\
&+F_{\sn}\frac{10}{3}i\mathcal{G}+F_{\se}\frac{10}{3}\mathcal{G}\cos{\iota}\,, \\
C_{\LiL}&=F_{+}\frac{\sin^2{\iota}}{2}+F_{\bb}\left(1-\frac{\sin^2{\iota}}{2}\right)\nn \\
&+F_{\se}\frac{\sin{2\iota}}{2}-F_{\LL}(3+\cos^2{\iota})\,.
\end{align}
In the derivation of the above equations we have used Kepler's law for this theory, which is identical to that of GR. 

Before we can compute the Fourier transform of this response function in the SPA, we must first find the orbital frequency evolution equation. The energy evolution of the binary's orbit is prescribed (to leading-order in the PN approximation and to leading-order in ${\cal{G}}$) by~\cite{Will:1977wq}
\be
\frac{dE}{dt}=\frac{84}{15}\frac{\mu^2 m^3}{r^5}+\frac{125}{9}\frac{\mu^2 m^2 \mathcal{G}^2}{r^4}\,.
\ee
Then, the orbital frequency evolution equation is simply
\be
\label{llfrequency}
\frac{dF}{dt}=-\frac{42}{5\pi\mathcal{M}^2}(2\pi\mathcal{M}F)^{11/3}-\frac{125}{6 \pi {\cal{M}}^{2}}\mathcal{G}^2 \eta^{2/5}  (2 \pi {\cal{M}} F)^3\,.
\ee

The SPA of the Fourier transform of the response function is 
\be
\label{eq:h-totalLL}
\tilde{h}_{\LiL}(f) = \tilde{h}_{\LiL}^{(1)}(f) + \tilde{h}_{\LiL}^{(2)}(f)\,, 
\ee
where the Fourier transform of the first term in Eq.~\eqref{llresponse} is 
\be
\tilde{h}_{\LiL}^{(2)}(f)=A_{\LiL}\;i\sqrt{\frac{5 \pi}{84}}\frac{\mathcal{M}^{2}}{D}(\pi \mathcal{M}f)^{-7/6}e^{-i\Psi_{\LiL}^{(2)}}\,,
\ee
and that of the second term is
\be
\tilde{h}_{\LiL}^{(1)}(f)=B_{\LiL}\;i\sqrt{\frac{5 \pi}{336}}\eta^{1/5} \frac{\mathcal{M}^{2}}{D}(\pi \mathcal{M}f)^{-3/2}e^{-i\Psi_{\LiL}^{(1)}}\,.
\ee
As before, we have here kept terms to leading-order in the PN approximation and in ${\cal{G}}$. We have also defined the $\ell^{\rm th}$-harmonic Lightman-Lee Fourier phase
\be
\label{Psi-LL}
\Psi_{\LiL}^{(\ell)}=\frac{\pi}{4}+\ell \Phi_c-2\pi f t_c+\frac{3 \ell }{224 u_{\ell}^5}-\frac{625 \ell}{16464}\frac{\mathcal{G}^2\eta^{2/5}}{u_{\ell}^7}\,,
\ee
where we recall that $u_{\ell}$ is given in Eq.~\eqref{eq:ul}. As in the case of Rosen's theory, note that Lightman-Lee theory does not have a well-defined GR limit, ie.~$\tilde{h}_{\LL}(f) \not \to \tilde{h}_{\GR}(f)$ as ${\cal{G}} \to 0$. Also note that some modifications are degenerate with system parameters, as we will see in Sec.~\ref{sec:ppEGen}. 

%---------------------------------------------------------------------------------------------
\section{Deconstruction of Inspiral Waveform Generation}
\label{sec:deconstruct}

With the information gathered so far, let us try to understand how the Fourier transform of the response function is constructed, allowing for possible modifications in the generation of GWs. The response function is clearly a sum of different terms in a harmonic decomposition, where the $\ell^{\rm th}$ term is 
\be
h^{(\ell)}(t) = Q(\iota,\theta,\phi,\psi) \; \eta^{2/5} \;  \frac{\cal{M}}{D} v^{\ell} e^{-i \ell \Phi}\,,
\label{eq:gen-form}
\ee
and where $Q(\iota,\theta,\phi,\psi)$ is a function of the inclination angle $\iota$ and possibly all the beam-pattern functions, which depend on the position of the source in the sky, parameterized by the angles $(\theta,\phi)$ and the polarization angle $\psi$. The $\ell=0$ harmonic does not satisfy this scaling, but as we have discussed in the previous sections, this mode contributes only subdominantly to the Fourier transform, and thus, it will be neglected henceforth.

Let us now argue by symmetry and dimensional reasoning that the functional form in Eq.~\eqref{eq:gen-form} is unavoidable for the $\ell$th harmonic. If one is to have a spacetime that is asymptotically flat, then the metric perturbation must scale as $D^{-1}$ to leading order. This quantity, however, has dimensions of inverse length, and thus, it must be accompanied by a quantity with dimensions of length. One could use $m_{1}$, $m_{2}$ or a combination like the reduced mass $\mu$. But since we must have that in the limit $m_{1} \to 0$ or $m_{2} \to 0$, then $h \to 0$, one cannot normalize $D$ with $m_{1}$ or $m_{2}$, or for the same reason with $\delta m = m_{1} - m_{2}$. Thus, the only dimension-1 quantity available for a non-spinning binary is $\mu$, which preserves exchange symmetry and leads to a factor of $\eta^{2/5} {\cal{M}}$ in the numerator of Eq.~\eqref{eq:gen-form}. Also, since we will require that any modified gravity theory has a well-defined continuous GR limit in the weak field, we must disallow any other arbitrary $\eta^{d}$ dependence. Finally, a term that scales with the $\ell^{\rm th}$ harmonic of the orbital phase must be proportional to the velocity to the $\ell^{\rm th}$power.

We should note that in principle Eq.~\eqref{eq:gen-form} should be multiplied by an infinity series in velocity of the form $\sum_{n=0} \left(a_{\ell,n} + b_{\ell,n} \ln{v}\right) v^{n}$ to account for higher-order PN corrections. In GR, one finds that for the $\ell=1$ mode, $a_{1,0} = 0 = b_{1,0}$ and $a_{1,1}=0=b_{1,1}$, and thus the series starts at 1PN order with $a_{1,2} \neq 0 = b_{1,2}$.  In modified gravity theories, however, this suppression of the $\ell=1$ mode need not be present. In fact, in theories with a scalar breathing mode, one finds that an $\ell=1$ mode is usually excited at leading Newtonian order. This his why we have not included the series dependence in Eq.~\eqref{eq:gen-form}.

Given this generic time-domain response function, we wish to compute the Fourier transform in the SPA. For this, we need both the binding energy and the rate of change of this energy. The former can be parameterized as follows
\be
\label{eeq}
E = E_{\GR} \left[1 + A \left(\frac{m}{r}\right)^{p} \right]\,,
\ee
where we assume $A$ is small, such that the correction represents a small deformation away from GR. Such a binding energy modifies Kepler's third law as
\be
\omega^{2} = \frac{m}{r^{3}} \left[1 + \frac{1}{2} A p \left(\frac{m}{r} \right)^p\right]\,,
\ee
or in terms of its inverse
\be
r = \left(\frac{m}{\omega^{2}}\right)^{1/3} \left[1 + \frac{1}{6} A p \left(m \omega\right)^{2p/3} \right]\,.
\ee
Using $v \equiv r \omega$, the Virial theorem is then modified to
\be
v = (m \omega)^{1/3} \left[1 + \frac{1}{6} A p \left(m \omega\right)^{2p/3} \right]\,.
\ee
Thus, the binding energy to leading PN order becomes
\begin{align}
E &= -\frac{1}{2} \eta^{-2/5} \left(2 \pi {\cal{M}} F\right)^{2/3} \left[1 
\right. 
\nn \\
&\left.- \frac{1}{3} A \left(5 p - 6\right) \eta^{-2p/5} \left(2 \pi {\cal{M}} F\right)^{2p/3}\right]\,,
\end{align}
and the response function is
\begin{align}
h^{(\ell)}(t) &= Q(\iota,\theta,\phi,\psi) \; \eta^{(2-\ell)/5} \;  \frac{\cal{M}}{D} \left(2 \pi {\cal{M}} F\right)^{\ell/3} e^{-i \ell \Phi}
\nn \\
&\times
\left[1 + \frac{1}{6} A \; p \; \ell \; \eta^{-2p/5} \left(2 \pi {\cal{M}} F\right)^{2p/3}\right]\,.
\end{align}
We see that this agrees with the GR and the Brans-Dicke results for the time-domain response function, when written in terms of the orbital frequency. 

The next ingredient we need is the rate of change of the binding energy. To model this one usually invokes the balance law, by which the rate of change of the binding energy must be exactly balanced by the energy flux carried away from the system (out to spatial infinity or into event horizons) by propagating degrees of freedom. Assuming that the metric perturbation accepts a multipolar decomposition, the energy flux is going to be the sum of the square of the $(\ell+1)$ time-derivative of the $\ell^{\rm th}$ multipole. Let us assume a modification to the rate of change of the binding energy of the form
\be
\label{edoteq}
\dot{E} = \dot{E}_{\GR} \left[1 + B \left(\frac{m}{r}\right)^{q}\right]\,,
\ee
where $\dot{E}_{\GR}$ is the GR energy flux and the second term is assumed small relative to the first, as we are interested in small deformations away from GR. Using the modified Kepler's law, we can rewrite this as
\begin{align}
\dot{E} &= -\frac{32}{5} \left(2 \pi {\cal{M}} F\right)^{10/3} 
\left[1 + B \eta^{-2q/5} \left(2 \pi {\cal{M}} F\right)^{2 q/3} 
\right. 
\nn \\
&-\left.
\frac{1}{3} A \; p \; \eta^{-2p/5} \left(2 \pi {\cal{M}} F\right)^{2p/3} \right]\,.
\end{align}
We see here that there are two modifications to $\dot{E}$: one coming from the modification to the Kepler law acting on $\dot{E}_{\GR}$ and one coming directly from the modification to $\dot{E}$ with the GR expression for Kepler's law. The values of $p$ and $q$ determine which of these two modifications dominates. We have here neglected the non-linear term generated by the product of terms proportional to $A$ and $B$. 

We now have all the ingredients to compute the Fourier transform of $h^{(\ell)}$ in the SPA. For this, it is convenient to first compute the rate of change of the orbital frequency via the chain rule:
\begin{align}
\frac{dF}{dt} &= \frac{48}{5\pi{\cal{M}}^{2}}  \left(2 \pi {\cal{M}} F\right)^{11/3} \left[1 
\right.
\nn \\
&+\left. 
B \eta^{-2q/5} \left(2 \pi {\cal{M}} F\right)^{2q/3} 
\right.
\nn \\
&+\left. 
\frac{1}{3} A  \left(5 p^{2} -2 p - 6\right) \eta^{-2p/5} \left(2 \pi {\cal{M}} F\right)^{2p/3}\right]\,,
\end{align}
where again we have kept terms to leading-order in the PN approximation and in the deformation parameters. The Fourier transform of the response function in the SPA is then
\begin{align}
\tilde{h}^{(\ell)} &= \sqrt{\frac{5 \pi}{48 \ell}} Q \frac{{\cal{M}}^{2}}{D} \eta^{(2 - \ell)/5} u_{\ell}^{(2 \ell -11)/2} e^{-i \Psi^{(\ell)}}
\nn \\
&\times
\left[1 - \frac{1}{2} B \eta^{-2q/5} u_{\ell}^{2q} 
\right.
\nn \\
&+ \left. 
\frac{1}{6} A  \left(\ell p + 6 + 2 p - 5 p^{2}\right) \; \eta^{-2p/5} \; u_{\ell}^{2 p} \right]\,,
\label{hofell}
\end{align}
where $u_{\ell}$ was defined in Eq.~\eqref{eq:ul}. We see then that the modification to the energy flux introduces corrections to the amplitude of $q$PN order, while the modification to the binding energy introduces corrections of $p$PN order. 

This calculation shows explicitly that the leading-order amplitude correction induced by an $\ell=1$ mode will not be degenerate with PN amplitude corrections since the former enters at Newtonian order. That is, the $\ell=1$ amplitude correction in Eq.~\eqref{hofell} is proportional to $u_{1}^{-9/2}$, which corresponds to a $-1$PN order term relative to the GR Newtonian amplitude that scales as $u^{-7/2}$. This scaling is very different from the GR amplitude corrections, which in Eq.~\eqref{high-ell-modes-in-GR} we saw scale as $u_{1}^{-5/2}$, a $+1$PN order correction relative to the GR Newtonian amplitude. Such different frequency scalings suggest that modified gravity amplitude corrections induced by the $\ell=1$ mode will be weakly correlated with GR amplitude corrections.

The Fourier phase can be similarly computed from Eq.~\eqref{spa phase general}. We find $\Psi^{(\ell)} = \Psi_{\GR}^{(\ell)} + \delta\Psi^{(\ell)}$, where $\Psi_{\GR}^{(\ell)}$ was given in Eq.~\eqref{eq:PsiGR} and we have defined
\begin{align}
\delta\Psi^{(\ell)} &= \frac{5}{64} A \; \frac{\ell \left(5 p^{2} - 2 p - 6\right)}{(4-p)(5-2p)} \eta^{-2 p/5} u_{\ell}^{2p-5} 
\nn \\
&+
\frac{15}{64} B \frac{\ell}{(4-q)(5-2q)} \eta^{-2q/5} u_{\ell}^{2q-5}\,.
\label{deltapsi}
\end{align}
Again, the modification to the binding energy introduces modifications of $p$PN order, while the modification to the flux is of $q$PN order. 

The above expression clearly contains poles at certain values of $q$ and $p$. These are not physical, but just a consequence of the assumptions made when carrying out the integral in Eq.~\eqref{spa phase general}. When $p$ or $q$ equal $4$ or $5/2$, the integrand that defines the Fourier phase becomes inversely proportional to $F$, and thus, the result should be a logarithm, instead of a power law. For these special cases, we find
\begin{align}\label{poles}
\delta\Psi^{(\ell)}_{p=4,B=0} &= \frac{880}{9} A \eta^{-8/5} u^8_{\ell} \left[1- 3 \ln{u_{\ell}} + \delta\phi_{0}\right]\,,
\\
\delta\Psi^{(\ell)}_{A=0,q=4} &= \frac{40}{9} B \eta^{-8/5} u^8_{\ell} \left[1- 3 \ln{u_{\ell}} + \delta\phi_{0}\right]\,,
\\
\delta\Psi^{(\ell)}_{p=5/2,B=0} &= 30 A \eta^{-1} u^5_{\ell} \left[1+ 3 \ln{u_{\ell}} + \delta\phi_{0}\right]\,,
\\
\delta\Psi^{(\ell)}_{A=0,q=5/2} &= \frac{40}{9} B \eta^{-1} u^5_{\ell} \left[1 + 3 \ln{u_{\ell}} + \delta\phi_{0}\right]\,,
\end{align}
where $\delta \phi_{0}$ is an overall constant of integration. Since these are $6.5$PN order and $5$PN order corrections relative to the Newtonian GR term, one usually ignores them. 

A note of caution is due at this junction. Equation~\eqref{deltapsi} depends only on integer powers of the reduced frequency $u_{\ell}$. This is a direct consequence of our parameterization of the energy flux and the binding energy in Eqs.~\eqref{edoteq} and~\eqref{eeq}, respectively. In principle, these equations could be modified by other functions that cannot be represented as a power series about zero velocity. A typical example would be logarithmic terms that, for example, enter at high PN order in General Relativity due to non-linear propagation effects (see eg.~\cite{Blanchet:2002av} and references therein). Moreover, certain modified gravity theories could introduce {\emph{screened}} modifications, ie.~corrections that become ``active'' only above a certain frequency. Such effects would need to be modeled through a Heaviside function, which cannot be represented as a power-law. This, for example, is the case when dealing with massive Brans-Dicke gravity~\cite{Detweiler:1980uk,Cardoso:2011xi,Alsing:2011er,Yunes:2011aa}.

%%%%%%%%%%%%%%%%%%%%%%%%%%%%%%%%%%%%%%%%%%%%%%%%%%
\section{ppE Generalization}
\label{sec:ppEGen}

We here review the basics of standard ppE theory and extend this to allow for the direct presence of additional polarizations. We also show that for single-detector observations, the standard ppE framework is sufficient for the $\ell=2$ harmonic, but should be generalized to account for the $\ell=1$ harmonic component. We then generalize the ppE scheme to allow for multiple detectors. 

%------------------------------------------------------------
\subsection{Standard ppE Framework}
\label{sec:standardppE}

In the standard ppE framework, one considers possible deformations of the two GR polarizations ($h_{+}$ $h_{\times}$) as induced by corrections to the frequency evolution equation only. As we have seen, this will allow only for modifications in the $\ell=2$ harmonic of the Fourier transform. Such corrections can arise due to deformations to the binding energy of the binary or the energy flux carried by all degrees of freedom away from the binary. In Sec.~\ref{sec:deconstruct}, we computed such corrections to leading PN order and to leading-order in the deformation parameters. As Yunes and Pretorius found~\cite{Yunes:2009ke}, the corrections to the Fourier transform of the response function can be generically cast as
\be
\label{ppe}
\tilde{h}_{\ppE,0}(f)= {\cal{A}} \; u_2^{-7/2} e^{-i \Psi^{(2)}_{\GR}} (1+\alpha u_{2}^a)e^{i 2 \beta u_{2}^b}\,,
\ee
where we have inserted a factor of $2$ in the phase correction for future convenience. If the modifications to the binding energy enter at the same PN order as the modifications to the energy flux, ie.~$q=k=p$, then
\begin{align}
\alpha &= \left[\frac{1}{6} A \left(  6 + 4 k - 5 k^{2}\right) - \frac{1}{2} B\right] \eta^{-2k/5}\,,
\\
\beta &= -\frac{5}{64} \frac{1}{(4-k)(5-2k)} \left[A \left(5k^{2} - 2 k - 6\right) + 3 B \right]\eta^{-2k/5}\,,
\\
a &= 2 k\,,
\qquad
b = 2 k - 5\,.
\end{align}
If $p<q$, then to leading-order one recovers again Eq.~\eqref{ppe} with 
\begin{align}
\alpha &= \frac{1}{6} A \left( 6 + 4p  - 5 p^{2}\right) \eta^{-2p/5}\,,
\\
\beta &= -\frac{5}{64} \frac{1}{(4-p)(5-2p)} A \left(5p^{2} - 2 p - 6\right) \eta^{-2p/5}\,,
\\
a &= 2 p\,,
\qquad
b = 2 p - 5\,.
\end{align}
while if $q<p$, then you have Eq.~\eqref{ppe} with
\begin{align}
\alpha &= - \frac{1}{2} B \eta^{-2q/5}\,,
\\
\beta &= -\frac{15}{64} \frac{1}{(4-q)(5-2q)} B \eta^{-2q/5}\,.
\\
a &= 2 q\,,
\qquad
b = 2 q - 5\,.
\end{align}
We see then clearly that all possibilities can be mapped to the ppE waveform family of Eq.~\eqref{ppe}. 

The ppE waveform family of Eq.~\eqref{ppe} depends on the standard $5$ system parameters $\vec{\lambda}_{\GR} = ({\cal{A}},{\cal{M}},\eta,t_{c},\Phi_{c}^{(2)})$ and on $4$ ppE theory parameters $\vec{\lambda}_{\ppE,0} = (\alpha,a, \beta,b)$. The ppE parameters $(a,b)$ are pure numbers that take discrete values and characterize the type of physical modification to the GR prediction. The parameters $(\alpha,\beta)$ depend on the coupling constants of the modified gravity theory, as well possibly on the masses and spins of the binary, and they are the quantities we wish to constrain with observations. When $(\alpha,\beta) = 0$, one recovers GR, while for other values of ppE parameters, one recovers the predictions of other theories for the $\ell = 2$ harmonic. For example, the Brans-Dicke prediction for the $\ell=2$ harmonic is recovered when
\begin{align}
\label{beta-BD}
\alpha_{\BD} &= \frac{224}{3} \beta_{\BD}\,,
\qquad
\beta_{\BD} = -\frac{5}{7168} \eta^{2/5} S^{2} \xi\,,
\\
a_{\BD} &= -2\,,
\qquad
b_{\BD} = -7\,.
\end{align}
Rosen's and Lightman-Lee theory cannot be easily mapped because they do not possess a GR limit, and thus, they are not technically small deformations away from Einstein's theory (see Sec.~\ref{sec:MG}).

From a physical standpoint, Eq.~\eqref{ppe} makes perfect sense: the waveform depends on an amplitude and a phase degree of freedom, so any deformation of it can be mapped to a change in these two quantities. One could in principle Taylor expand these functional degrees of freedom, but lacking a particular theoretical framework, the controlling factor of such an expansion is unknown. Thus, since different theories predict different controlling factors, it is best to use the parameterization in Eq.~\eqref{ppe}. The pure numbers $(a,b)$, however, should not be thought of as completely free, as after all they are generated by velocity corrections to GR's physical principles. If one assumes that modified gravity theories always lead to corrections of the form $v^{n}$, where $n \in \mathbb{Z}$, then $(a,b) \in \mathbb{Z}$. To date, we have not encountered any modified gravity theory that predicts corrections in non-integer powers of the orbital velocity. 

Can modifications to the radiation-reaction force or the Hamiltonian in modified gravity theories depend on non-integer powers of the velocity? We believe the answer to this question is no. If such terms were present, the corrections would be non-analytic in the weak-field, ie.~they would lack a well-defined Taylor expansion about zero velocity. Such lack of analyticity would lead to a breakdown of the initial value formulation. For example, differential equations of the form $\dot{v} = v^{p}$, with $p$ a non-integer and initial condition $v(0) = 0$ do not possess a unique solution. Notice that this is not a problem for terms proportional to negative powers of velocity or to natural logarithms, as the latter are analytic everywhere in their domain $(0,\infty)$, ie.~$v=0$ is not in the domain of these functions. 

However, it is known that non integer powers of the frequency in the phase of the GWs arise when the orbit is no longer circular. The reason the mathematical argument presented above is no longer valid is the following. For eccentric orbits, the binding energy is a bivariate series both in the azimuthal velocity $u_{\theta}$ and in the radial velocity $u_{r}$. The former is initially non-zero, but very close to zero, increasing as the inspiral proceeds. Therefore, one can expand the binding energy around $u_{\theta}=0$ and a non-integer power in the evolution law would break uniqueness. On the other hand, during the inspiral $u_{r}$ decays. An inspiral can begin with large eccentricity, but due to radiation reaction, it would lose eccentricity and become circular as the inspiral proceeds. Therefore, one can no longer expand the binding energy around $u_{r}=0$, which implies the initial condition $u_{r}(0) = 0$ is no longer valid. A differential equation $\dot{u}_{r} = u_{r}^{p}$ with $p$ non-integer and an initial condition $u_{r}(0) \ne 0$ has no uniqueness problem. Therefore, if an extra field $\phi$ were present and it decayed during the inspiral without an initial condition $\phi(0) = 0$, one cannot use the argument above to rule out non-integer powers in the evolution of the GW phase. 

The parameterization of Eq.~\eqref{ppe} does not use the fact that a modification to GW generation will usually introduce changes to the amplitude and phase that will be related to each other. Indeed, from the analysis of Sec.~\ref{sec:deconstruct}, we see that a parameterization of the $\ell=2$ harmonic of the Fourier response that explicitly recognizes these relations would be
\be
\tilde{h}_{\ppE,1} = {\cal{A}} \; u_2^{-7/2} e^{-i \Psi^{(2)}_{\GR}} \left[1 + c \; \beta \; u_{2}^{b+5} \right] 
e^{i 2 \beta \, u_{2}^{b}}\,.
\label{eq:simple-ppE}
\ee
Notice that $a$ and $\alpha$ have been completely eliminated. The parameter $c \in \mathbb{R}$ is not a free parameter, but it is fully determined by $b$. When a conservative correction dominates, then 
\be
\label{c-cons}
c_{\rm{cons}} = -\frac{16}{15} \frac{b(3 - b) (42 b + 61 + 5 b^{2})}{5 b^{2} + 46 b + 81}\,,
\ee
when a dissipative correction dominates, then
\be
\label{c-diss}
c_{\rm{diss}} = -\frac{16}{15} (3 - b) b\,,
\ee
and if both conservative and dissipative corrections enter at the same PN order, then
\be
c_{\rm{both}} =- \frac{32}{15} \frac{b(3 -b) (44b +71 + 5 b^{2})}{5b^{2} + 46 b + 81}\,.
\label{c-both}
\ee
Therefore, the ppE theory parameters of $\tilde{h}_{\ppE,1}$ are $\vec{\lambda}_{\ppE,1} = (b,\beta)$, two less than $\vec{\lambda}_{\ppE,0}$. This might seem strange since originally we started with 4 arbitrary constants $(A,B,p,q)$, two to parameterize corrections to the binding energy and two for modifications to the flux. Without loss of generality, however, either one of these two will be dominant or they will enter at the same PN order, enabling us to eliminate $2$ parameters. Notice that when $b = -7$ and a dissipative correction dominates, then $c_{\rm{diss}} = 224/3$ and one recovers the Brans-Dicke result of Eq.~\eqref{beta-BD}.

This reduced parameterization, however, neglects possible modifications in the propagation of GWs, due for example to a Lorentz-violating graviton dispersion relation. The analysis of~\cite{Will:1997bb,Mirshekari:2011yq} showed that modifications in GW propagation only introduce corrections to the Fourier phase, and thus, Eq.~\eqref{eq:simple-ppE} should be modified to 
\be
\tilde{h}_{\ppE,2} = {\cal{A}} \; u_2^{-7/2} e^{-i \Psi^{(2)}_{\GR}} \left[1 + c \; \beta \; u_{2}^{b+5} \right] 
e^{i 2 \beta \, u_{2}^{b}} e^{i \kappa u_{2}^{k}}\,,
\ee
where $k \in \mathbb{Z}$ while $\kappa \in \mathbb{R}$ depends on the coupling constants of the theory. Obviously, if one neglects the $2 \beta u_{2}^{b}$ term in the phase, one recovers exactly $\tilde{h}_{\ppE,0}$ after an appropriate relabeling of the ppE parameters [see Eq.~\eqref{ppe}]. This is only valid if $k<b$, as then the $2 \beta u_{2}^{b}$ term would be of higher PN order than the $\kappa u_{2}^{k}$ term. This template family has parameters $\vec{\lambda}_{\ppE,2} = (b,\beta,\kappa,k)$, the same number as $\vec{\lambda}_{\ppE,0}$, although there is a clear phase degeneracy here when $k = b$. The ppE template family $\tilde{h}_{\ppE,2}$ allows one to disentangle generation from propagation effects.

The parameterizations discussed above, however, only work for a single-detector and for the $\ell=2$ harmonic. In the next subsections, we generalize these proposals to allow for power in the $\ell=1$ harmonic and for the possibility of multiple-detector detections. 

%------------------------------------------------------------
\subsection{Generalized ppE Scheme: \\ Single Detector, all harmonics}
\label{sec:gen-sd}

Given a single interferometric detector, one cannot break the degeneracies in the waveform to separately extract all physical parameters~\cite{Cornish:2011ys}. For example, one cannot separately measure all the angles that characterize the position of the source in the sky. Thus, even within GR, the prototypical template family is a reduction of Eq.~\eqref{eq:GR-SPA} to
\be
\tilde{h}^{\GR}_{\SD}(f)= {\cal{A}}_{\GR} (\pi \mathcal{M} f)^{-7/6} \; e^{-i\Psi_{\GR}}\,, 
\ee
where we have absorbed a constant factor into the phase of coalescence. That is, one can measure the parameters contained in $\Psi_{\GR}$, ie.~$(t_{c},\Phi_{c},\eta,{\cal{M}})$, plus the overall amplitude ${\cal{A}}_{\GR}$, which essentially determines the signal-to-noise ratio. However, one cannot separately measure all the quantities encoded in
\be
{\cal{A}}_{\GR} = \left(\frac{5 \pi}{96}\right)^{1/2} \frac{{\cal{M}}^{2}}{D} \left[F_{+}^{2} \left(1 + \cos^{2}{\iota}\right)^{2} + 4 F_{\times}^{2} \cos^{2}{\iota} \right]^{1/2}\,. 
\ee

Similarly, when deciding how to modify the GR template family to allow for modified gravity effects, one must keep possible degeneracies in mind. Below, we calculate the reductions of the SPA Fourier transform of the response functions of Sec.~\ref{sec:MG} for each modified gravity theory studied and then discuss a few ppE proposals.  

%------------------------
\subsubsection{Brans-Dicke Theory} 
Equation~\eqref{eq:full-BD} can be rewritten as
\begin{align}\label{bdwhole}
\tilde{h}^{\BD}_{\SD}(f) &= {\cal{A}}_{\BD} u_2^{-7/2} \left[1+ \frac{224}{3} \beta_{\BD} u_{2}^{-2} \right] e^{-i \Psi_{\BD}^{(2)}}
\nn \\
&+ \gamma_{\BD} u_{1}^{-9/2} e^{-i \Psi_{\BD}^{(1)}}\,.
\end{align}
where $\Psi_{\BD}^{(\ell)} = \Psi_{\GR}^{(\ell)} + \delta \Psi_{\BD}^{(\ell)}$ is given by Eq.~\eqref{psi1}, with the substitution $\Phi_{c} \to \Phi_{c}^{(\ell)}$ in $\Psi_{\GR}^{(\ell)}$. The Brans-Dicke correction can be written as $\delta\Psi_{\BD}^{(\ell)} = -({\ell}/{2}) \beta_{\BD} u_{\ell}^{-7}$, where $\beta_{\BD}$ was already defined in Eq.~\eqref{beta-BD} and we have redefined the chirp mass $\bar{{\cal{M}}} =  {\cal{M}}[1 - (\xi/15)(\Gamma^{2}/12 - k_{\BD})]$. In rewriting the waveform in Eq.~\eqref{eq:full-BD} in terms of an overall amplitude and an overall phase, we also redefined the (constant) phase of coalescence  as 
\begin{align}
\Phi_{c}^{(\ell)} &= \Phi_{c} - \delta_{\ell,2}\left\{\arctan{\left[\frac{2 \cos{\iota} F_{\times}}{(1+\cos^2{\iota})F_{+}}\right]}\right.\nn\\
&\left.+\frac{\xi \Gamma\cos{\iota}(1-\cos^2{\iota})F_{\times}F_{\bb}}{(1+\cos^2{\iota})^2F_{+}^2+4\cos^2{\iota}F_{\times}^2}\right\}\,,
\end{align}
which leads to two distinct constants for each harmonic. Therefore, the parameters of such a template family are the usual system parameters $\vec{\lambda}_{\GR}$, plus the Brans-Dicke parameters $\vec{\lambda}_{\BD} = (\beta_{\BD},\gamma_{\BD},\Phi_{c}^{(1)})$. 

For a single-detector, one still detects only a combined overall amplitude ${\cal{A}}_{\BD}$ for the $\ell = 2$ harmonic, which in terms of the fundamental parameters of the theory and the beam-pattern functions is given by
\begin{align}
{\cal{A}}_{\BD} &= \sqrt{\frac{5\pi}{96}} \frac{{\cal{M}}^{2}}{\bar{D}} 
\left[ 
F_{+}^{2} \left(1 + \cos^{2}{\iota}\right)^{2}  + 4 F_{\times}^{2} \cos^{2}{\iota}
\right. 
\nn \\
&+ \left.
F_{+} F_{\bb} \left( \cos^{4}{\iota}-1 \right) \; \xi \; \Gamma
\right]^{1/2}\,,
\end{align}
and we recall that $\Gamma$ is defined below Eq.~\eqref{S-def}. We have here absorbed a prefactor into the luminosity distance, namely $\bar{D} = D \left[1 +  \xi  \left(\Gamma^{2}/24 + k_{\BD}/2\right)\right]$, where $k_{\BD}$ and $S$ are defined in and around Eq.~\eqref{S-def}. Moreover, for a single-detector one cannot measure $\xi$ separately, but the combined phase and amplitude quantities. The latter is related to $\xi$ via 
\begin{align}
\gamma_{\BD} &= - \left(\frac{5 \pi}{48} \right)^{1/2}
 \frac{{\cal{M}}^{2}}{D} \eta^{1/5}
S \; \xi \; F_{\bb} \sin{\iota}\,,
%\\
%\beta_{\BD} &=- \frac{5}{7168} \eta^{2/5} \; S^{2} \; \xi\,,
\end{align}
We separate $\gamma_{\BD}$ from $\beta_{\BD}$ here because they appear multiplied by different powers of $u_{\ell}$ and the beam-pattern functions in the Fourier transform of the response function. That is, different (single) detectors will measure the same $\beta_{\BD}$, but different $\gamma_{\BD}$.  

%------------------------
\subsubsection{Rosen's Theory} 

Equation~\eqref{eq:full-h-Rosen} can be rewritten as
\begin{align}
\tilde{h}^{\R}_{\SD}(f)&={\cal{A}}_{\R} u_2^{-7/2}e^{-i\Psi_{\R}^{(2)}}+\gamma_{\R}u_1^{-9/2}e^{-i\Psi_{\R}^{(1)}}\,,
\end{align}
where $\Psi_{\R}^{(\ell)}$ is given by Eq.~\eqref{eq:Rosen-Psi1} with the substitution $\Phi_{c} \to \Phi_{c}^{(\ell)}$. The last term in Eq.~\eqref{eq:Rosen-Psi1} can also be written as $\beta_{\R} u_{\ell}^{-7}$, where $\beta_{\R}$ is given below.  As in the Brans-Dicke case, the above expression is obtained by rewriting Eq.~\eqref{eq:full-h-Rosen} in terms of an overall amplitude and phase, where we have redefined the (constant) phase of coalescence via
\begin{align}
\Phi_c^{(\ell)} &= \Phi_{c}- \delta_{\ell,1} \arctan\left(-\frac{F_{\bb} \sin{\iota} + F_{\LL} \sin{\iota} + F_{\se} \cos{\iota}}{F_{\sn}} \right)\nn \\
&- \delta_{\ell,2} \arctan\left[ -\frac{F_{+}(1+\cos^2{\iota})}{2F_{\times}\cos{\iota}-2F_{\sn}\sin{\iota}}
\right.
\nn \\
&-\left. \frac{(F_{\bb}+F_{\LL})\sin^2{\iota}+F_{\se}\sin{2\iota}}{2F_{\times}\cos{\iota}-2F_{\sn}\sin{\iota}} \right]\,.
\end{align}
Therefore, the parameters of this template family are the usual system parameters $\vec{\lambda}_{\GR}$, plus the Rosen parameters $(\beta_{\R},\gamma_{\R},\Phi_{c}^{(1)})$. 

As before, a single detector can only measure the combined amplitude ${\cal{A}}_{\R}$, which is related to other system parameters via
\begin{align}
{\cal{A}}_{\R} &= \sqrt{\frac{5 \pi}{84}} \frac{\mathcal{M}^2 k_{\R}^{-3/4}}{4 D} 
\left[ F_{+}^{2} \left(1 + \cos^{2}{\iota}\right)^{2} + 4 F_{\times}^{2} \cos^{2}{\iota} 
\right. 
\nn \\
&+ \left.
F_{\bb} \sin^{4}{\iota} + 4 F_{\LL}^{2} \sin^{4}{\iota} + F_{\se}^{2} \sin^{2}{2\iota} + 4 F_{\sn}^{2} \sin^{2}{\iota} 
\right.
\nn \\
&+ \left. 
8 \cos{\iota} \sin^{3}{\iota} F_{\LL} F_{\se}
+4  \sin^{4}{\iota} F_{\bb} F_{\LL}
+4\cos{\iota} \sin^{3}{\iota} F_{\bb} F_{\se}
\right. 
\nn \\
&+\left. 
2 \left(1 - \cos^{4}{\iota}\right) \left(F_{+} F_{\bb} + 2 F_{+} F_{\LL}\right)
\right. 
\nn \\
&+\left. 
(1/2) \left(6 \sin{2\iota} + \sin{4\iota}\right) F_{+} F_{\se}
+ 4 \sin{2\iota} F_{\times} F_{\sn} \right]^{1/2}\,,
\end{align}
One can also measure the combination of theory constants in $\beta_{\R}$ and $\gamma_{\R}$, where the relationship between these two sets and theory parameters is
\begin{align}
\gamma_{\R} &= \sqrt{\frac{40 \pi}{189}} \eta^{1/5} \frac{\mathcal{M}^{2} k_{\R}^{-7/12}}{D} {\cal{G}} 
\left[F_{\sn}^{2} + \left(F_{\bb}^{2} + F_{\LL}^{2}\right) \sin^{2}{\iota} 
\right. 
\nn \\
&+\left.
F_{\se}^{2} \cos^{2}{\iota} +  2 F_{\bb} F_{\LL} \sin^{2}{\iota} + \left(F_{\bb} + F_{\LL}\right) F_{\se} \sin{2\iota}
\right]^{1/2}
\,,
\\
\beta_{\R} &=- \frac{25}{8232} {k_{\R}^{-2/3}\cal{G}}^{2} \eta^{2/5}\,. 
\end{align}
As in the Brans-Dicke case, all detectors will see the same $\beta_{\R}$ for a given event, but different $\gamma_{\R}$. 

%------------------------
\subsubsection{Lightman-Lee Theory} 

Equation~\eqref{eq:h-totalLL} can be rewritten as
\begin{align}
\tilde{h}^{\LiL}_{\SD}(f)={\cal{A}}_{\LiL} u_2^{-7/2}e^{-i\Psi_{\LiL}^{(2)}}+\gamma_{\LiL}u_1^{-9/2}e^{-i\Psi_{\LiL}^{(1)}}\,,
\end{align}
where $\Psi_{\LiL}^{(\ell)}$ in this equation is given by Eq.~\eqref{Psi-LL} with the substitution $\Phi_{c} \to \Phi_{c}^{(\ell)}$. The last term in Eq.~\eqref{Psi-LL} can also be written as $\beta_{\LiL} u_{\ell}^{-7}$, where $\beta_{\LiL}$ is given below. In rewriting Eq.~\eqref{eq:h-totalLL} in terms of an overall amplitude and phase, we had to redefine the constant phase of coalesce to 
\begin{align}
\Phi_{c}^{\ell} &= \Phi_{c} -\delta_{\ell,2} 
\arctan\left[-
\frac{F_{+}\left(1 + \cos^{2}{\iota}\right)}{2 F_{\sn} \sin{\iota}  + 2  F_{\times} \cos{\iota}}
\right.
\nn \\
&+ \left.
\frac{-\left( 2 F_{\LL}  - 3 F_{\bb}\right) \sin^{2}{\iota} + F_{\se} \sin{2 \iota}}{2 F_{\sn} \sin{\iota}  + 2  F_{\times} \cos{\iota}}
\right]
\nn \\
&- \delta_{\ell,1} \arctan\left\{\frac{\left[(5/4) F_{\bb} + F_{\LL}\right]\sin{\iota} - \cos{\iota} F_{\se}}{F_{\sn}}\right\}\,.
\end{align}
Therefore, the parameters of this waveform are the usual ones $\vec{\lambda}_{\GR}$, plus the Lightman-Lee ones $(\beta_{\LiL},\gamma_{\LiL},\Phi_{c}^{(1)})$. 

As before, a single detector can only measure the overall amplitude ${\cal{A}}_{\LiL}$, which is related to other parameters via 
\begin{align}
{\cal{A}}_{\LiL} &= \sqrt{\frac{5 \pi}{168}} \frac{\mathcal{M}^2}{D}
\left[ F_{+}^{2} \left(1 + \cos^{2}{\iota}\right)^{2} + 4 \cos^{2}{\iota} F_{\times}^{2} 
\right.
\nn \\
&+ \left.
 \left(9 F_{\bb}^{2} + 4 F_{\LL}^{2} \right)\sin^{4}{\iota} 
+\sin^{2}{2 \iota} F_{\se}^{2} + 4 \sin^{2}{\iota} F_{\sn}^{2}
\right.
\nn \\
&+ \left.
8 \cos{\iota} \sin^{3}{\iota} F_{\LL} F_{\se}
- 12 \sin^{4}{\iota} F_{\bb} F_{\LL} - 12 \sin^{3}{\iota} \cos{\iota} F_{\bb} F_{\se} 
\right.
\nn \\
&- \left.
(1/2) \left(5 \sin{\iota} + \sin{3\iota} \right)
F_{+}\left(3 \sin{\iota} F_{\bb} - 2 \sin{\iota} F_{\LL} 
\right.\right.
\nn \\
& \left.\left.
- 2 \cos{\iota} F_{\se}\right) + 4 \sin{2 \iota} F_{\sn} F_{\times} 
\right]^{1/2}
\,,
\end{align}
Moreover, one can also only measure $(\gamma_{\LiL},\beta_{\LiL})$, and these are related to other fundamental theory theory via
\begin{align}
\gamma_{\LiL} &=\sqrt{\frac{250\pi}{189}}\eta^{1/5} \frac{\mathcal{M}^{2}}{D} {\cal{G}} 
\left[\left(\frac{25}{16} F_{\bb}^{2} + F_{\LL}^{2} \right)\sin^{2}{\iota} 
\right. 
\nn \\
&+ \left.
\cos^{2}{\iota} F_{\se}^{2} + F_{\sn}^{2} + \frac{5}{2} \sin^{2}{\iota} F_{\bb} F_{\LL} 
\right. 
\nn \\
&- \left.
 F_{\LL} F_{\se} - \frac{5}{4} F_{\bb} F_{\se} \sin{2 \iota}  
\right]^{1/2}
\,,
\\
\beta_{\LiL} &=  \frac{625}{16464} {\cal{G}}^{2} \eta^{2/5}\,.
\end{align}
Multiple detectors will measure different $\gamma_{\LiL}$, but the same $\beta_{\LiL}$ given the same source.

%---------------------------------------
\subsubsection{ppE Scheme}

The above discussion allows us to conclude that, although the already existing ppE scheme is perfectly adequate to capture the $\ell=2$ harmonic in the Fourier transform, it does not capture the $\ell=1$ harmonic. Power in this harmonic is a natural consequence of theories that predict extra polarization states. The largest generalization of the ppE scheme that covers all alternatives is simply 
\begin{align}\label{ppE1}
\tilde{h}_{\ppE,0}^{\SD}(f)&={\cal{A}} \; u_{2}^{-7/2} (1+\alpha u_{2}^a)e^{-i \Psi_{\GR}^{(2)}}\; e^{i\beta u_{2}^b} 
\nn \\
&+ \gamma \; u_{1}^c \; e^{-i \Psi_{\GR}^{(1)}} \; e^{i\delta u_{1}^d}\,,
\end{align} 
where now the theory parameters are $\vec{\lambda}_{\ppE,0}=(\alpha,a,\beta,b,\gamma,c,\delta,d,\Phi_{c}^{(1)})$. We say that this is the largest generalization because, in addition to the standard $5$ system parameters $\vec{\lambda}_{\GR}$, we have added $9$ more for a total of $14$ parameters, without imposing any relations between them. 

One might worry that the parameterization of the $\ell=1$ term (the second line in Eq.~\eqref{ppE1} will be degenerate with PN amplitude corrections. This will indeed be the case if $c \in (-5/2,-2,-3/2,\ldots)$. Modified gravity theories that predict the excitation of a scalar breathing mode, however, are likely to introduce $\ell=1$ corrections to the waveform that enter at leading, Newtonian order. For these, $c = -9/2$, which is a $-1$PN order correction in the amplitude relative to the leading-order GR term. As such, these modifications should be weakly correlated with PN amplitude corrections.

Since the leading-order term of the $\ell=1$ harmonic enters at Newtonian, leading order, the amplitude parameter $\gamma$ has to be very small, for the theory to be consistent with binary pulsar observations. Given that detectors are more sensitive to the GW phase than its amplitude, it would appear that the effect of additional polarization states would first show up indirectly in the modified phase evolution of $h_+$ and $h_{\times}$, and not through the direct detection of the additional polarization states. However,  $\dot{E} \sim \dot{h}^2 \sim O(\epsilon^2)$, where $\epsilon$ is the small parameter measuring the deviation of the theory in question from GR. Then, it is clear from the SPA formalism (Eq.~\eqref{spa general}) that the amplitude of the Fourier transform of the $\ell=1$ harmonic will have 2 terms: one coming from the quadrupole part of $\dot{F}$ and one coming from the dipole part of $\dot{F}$. The former is of order $\epsilon$, while the latter of order $\epsilon^2$ and can, thus, be neglected. From Eq.~\eqref{spa phase general} we see that $\beta$ is proportional to $\epsilon^2$ and we still have the possibility of measuring $\gamma$ before $\beta$.  Only in very specific theories, for example in Brans-Dicke theory where the kinetic scalar term is multiplied by the reciprocal of the Brans-Dicke coupling parameter, is $\dot{E}$ modified from its GR expectation at linear order in the coupling. A more detailed study of the possibility of measuring $\gamma$ is left for future work.
 
In the above parameterization, we did not use the fact that some of these parameters are not independent, if they are all to arise from a modification to the generation of GWs. As we showed in Sec.~\ref{sec:deconstruct}, given a correction to the energy flux or the binding energy that scales as $1/r^{k}$ relative to the leading-order GR expectation, the correction to the phase in the $\ell=2$ harmonic is proportional to $u_{2}^{2k-5}$, while the correction to the amplitude goes as $u_{2}^{2k}$. One can then postulate a reduced ppE model of the form
\begin{align}\label{ppE2}
\tilde{h}^{\SD}_{\ppE,1}(f) &= {\cal{A}} \; u_2^{-7/2} e^{-i \Psi^{(2)}_{\GR}} \left[1 + c \; \beta u_{2}^{b+5} \right] 
e^{i 2 \beta u_{2}^{b}}
\nn \\
&+ \gamma \; u_{1}^{-9/2} e^{-i \Psi^{(1)}_{\GR}}
e^{i \beta u_{1}^{b}}\,,
\end{align}
where $\Psi_{\GR}^{(\ell)}$ is given in Eq.~\eqref{eq:PsiGR} with $\Phi_{c} \to \Phi_{c}^{(\ell)}$. We have also explicitly used the fact that the $\ell=2$ harmonic leads to an amplitude proportional to $f^{-7/6}$, while an $\ell=1$ harmonic leads to $f^{-3/2}$ when only the leading-order quadrupole radiation is taken into account. Notice that $c$ is again given by Eqs.~\eqref{c-cons}-~\eqref{c-both}, ie.~it is fully determined by the conservative or dissipative modifications. The ppE theory parameters of $\tilde{h}_{\ppE,1}^{\SD}$ are $\vec{\lambda}_{\ppE,1} = (b,\beta,\gamma,\Phi_{c}^{(1)})$, in addition to the usual system parameters $\vec{\lambda}_{\GR}$, for a total of $9$ parameters. 

The result in Eq.~\eqref{ppE2} is very similar to that of Arun~\cite{Arun:2012hf}, so let us compare it directly. In~\cite{Arun:2012hf}, Arun considered the effect of dipole radiation in the Fourier waveform, which corresponds to our waveform with $b = -7$. Moreover, he neglected any conservative corrections (modifications to the binding energy), and thus, $c = c_{\rm{diss}}$ as given in Eq.~\eqref{c-diss}. In this limit, $\tilde{h}_{\ppE,1}^{\SD}$ reduces to his Eq.~$(9)$. We note that Arun's overall amplitude is complex, and thus, there is an additional parameter that was not made explicit in his equations. We have here pulled this factor out by including $\Phi_{c}^{(1)} \neq \Phi_{c}^{(2)}$ explicitly. 

Whether one chooses Eq.~\eqref{ppE1} or~\eqref{ppE2} depends on the question that one is trying to answer. The parameterization in Eq.~\eqref{ppE2} contains fewer parameters, but stronger assumptions about the type of corrections one is searching for. In deriving this equation, we have neglected all corrections that cannot be represented by a power-law. For example, we have ignored possible logarithmic modified gravity corrections in the Hamiltonian or radiation-reaction force, as well as screened modifications that depend on the Heaviside function. Moreover, we have also neglected the fact that GR modifications consist of an infinite power series in velocity, and thus, corrections to the phase will not depend on a single term, but an entire PN series. The parameterization of Eq.~\eqref{ppE2} attempts to extract only the leading-order, controlling factor in the expansion. If the true theory of nature contains these type of deviations, Eq.~\eqref{ppE1} might be more powerful at detecting a deviation. This is the topic of a follow-up study that is in preparation~\cite{Laura:prep}.  

Finally, we can also allow for propagation effects by multiplying the phases in Eq.~\eqref{ppE2} by the appropriate ppE propagation factors:
\begin{align}\label{ppE3}
\tilde{h}^{\SD}_{\ppE,2}(f) &= {\cal{A}} \; u_2^{-7/2} e^{-i \Psi^{(2)}_{\GR}} \left[1 + c \; \beta u_{2}^{b+5} \right] 
e^{i 2 \beta u_{2}^{b}} e^{i \kappa u_{2}^{k}}
\nn \\
&+ \gamma \; u_{1}^{-9/2} e^{-i \Psi^{(1)}_{\GR}}
e^{i \beta u_{1}^{b}} e^{i \kappa u_{1}^{k}}\,,
\end{align}
where $(\kappa,k)$ are new ppE propagation parameters. Using such a waveform, however, might not be ideal for data analysis purposes, as usually either $u^{b}$ or $u^{k}$ will dominate, and it might be preferable to ignore the subdominant contribution.

%----------------------------------------------------------
\subsection{Generalized ppE Scheme: \\ Multiple Detectors, all harmonics}

The theory amplitude parameters, such as $\beta$ and $\gamma$, introduced in the previous subsection, clearly depend both on the coupling constants of the theory, as well as on the symmetric mass ratio, the sky location of the source and the inclination angle. Therefore, these parameters are system and detector dependent and not suitable for comparing data acquired from a detector network. In this subsection, we will construct a generalization to the ppE parameterization for the waveform amplitude to resolve this problem. 

Let us then begin by presenting the Fourier transform of the response function in the SPA for all the theories considered in Sec.~\ref{sec:MG}, but in full detail this time, without rewriting the response in terms of an overall amplitude and phase correction.
\begin{enumerate}
\item {\emph{GR}}:
\begin{align}
\label{grmany}
\tilde{h}^{\GR}_{\MD}(f) &=-[F_{+}(1+\cos^2{\iota})+2iF_{\times}\cos{\iota}]
\nonumber \\
&\times
\left(\frac{5\pi}{96}\right)^{1/2}\frac{\mathcal{M}^{2}}{D}(\pi \mathcal{M} f)^{-7/6}e^{-i\Psi_{\GR}^{(2)}}
\end{align}
\end{enumerate}
\begin{widetext}
\begin{enumerate}
\item[2.] {\emph{Brans-Dicke Theory}}:
\begin{align}\label{bdmany}
\tilde{h}^{\BD}_{\MD}(f)&=\left\{-[F_{+}(1+\cos^2{\iota})+2iF_{\times}\cos{\iota}]\left[1 - \left(\frac{1}{24}\Gamma^2+\frac{k_{\BD}}{2}\right) \xi \right] u_{2}^{-7/2}
+F_{\bb} \sin^2{\iota} \; \frac{\Gamma}{2} \; \xi \; u_{2}^{-7/2}
\right.
\nn\\
&\left.-[F_{+}(1+\cos^2{\iota})+2iF_{\times}\cos{\iota}]  \left(\frac{5}{96}\right)  \eta^{2/5} \; \xi \; S^2 \; u_{2}^{-11/2}
\right\}
\left(\frac{5\pi}{96}\right)^{1/2}\frac{\mathcal{M}^{2}}{D}e^{-i\Psi_{\BD}^{(2)}}
\nn\\
&-\left(F_{\bb}\; \sin{\iota} \right) S \; \xi 
\left(\frac{5\pi}{48}\right)^{1/2} \eta^{1/5} \frac{\mathcal{M}^{2}}{D} 
u_{1}^{-9/2} e^{-i\Psi_{\BD}^{(1)}}\,,
\end{align}
\end{enumerate}

\begin{enumerate}
\item[3.] {\emph{Rosen's Theory}}:
\begin{align}
\tilde{h}^{\R}_{\MD}(f)&= \left\{-\left[F_{+}(1+\cos^2{\iota})+2iF_{\times}\cos{\iota}\right]
\right.
\nn\\
&\left.+ \left[-F_{\bb}\sin^2{\iota}-2F_{\LL}\sin^2{\iota}-2F_{\sn}i\sin{\iota}-F_{\se}\sin{2\iota} \right]\right\}
i \left(\frac{5\pi}{336}\right)^{1/2} k_{\R}^{-3/4}\frac{\mathcal{M}^{2}}{D} u_{2}^{-7/2} e^{-i\Psi_{\R}^{(2)}}
\nn\\
&+i \left(-F_{\bb}\sin{\iota}-F_{\LL}\sin{\iota}-F_{\sn}i-F_{\se}\cos{\iota}\right)\mathcal{G} k_{\R}^{-7/12} \sqrt{\frac{40\pi}{189}} \eta^{1/5} \frac{\mathcal{M}^{2}}{D} u_{1}^{-9/2 }e^{-i\delta\Psi_{\R}^{(1)}}\,,
\end{align}
\end{enumerate}

\begin{enumerate}
\item[4.] {\emph{Lightman-Lee Theory}}:
\begin{align}
\tilde{h}^{\LiL}_{\MD}(f)&= \left\{-\left[F_{+}\left(1+\cos^2{\iota}\right)+2iF_{\times}\cos{\iota}\right]
+ \left[3F_{\bb}\sin^2{\iota}-2F_{\LL}\sin^2{\iota}-2F_{\sn}i\sin{\iota}-F_{\se}\sin{2\iota} \right]\right\}
\nn \\
&\times i \left(\frac{5\pi}{336}\right)^{1/2}\frac{\mathcal{M}^{2}}{D} u_{2}^{-7/2} e^{-i\Psi_{\LiL}^{(2)}}
+\left[\frac{5}{4}F_{\bb}\sin{\iota}-F_{\LL}\sin{\iota}+F_{\sn}+F_{\se}\cos{\iota}\right] i \; \mathcal{G} \; \sqrt{\frac{250 \pi}{189}} \eta^{1/5} \frac{\mathcal{M}^{2}}{D}
u_{1}^{-9/2}e^{-i\delta\Psi_{\LiL}^{(1)}}\,,
\end{align}
\end{enumerate}
where the phase terms are given in Sec.~\ref{sec:MG}.

Using the above examples for guidance, one can generalize the ppE framework to apply to multiple detectors via the large ppE class 

\begin{align}\label{ppE1md}
\tilde{h}_{\ppE,0}^{\MD}(f)&=\tilde{h}^{\GR}_{\MD} e^{i\beta u_{2}^b}+[\alpha_{+} F_{+} +\alpha_{\times} F_{\times} + \alpha_{\bb} F_{\bb}+\alpha_{\LL}F_{\LL}+\alpha_{\sn}F_{\sn}+\alpha_{\se}F_{\se}]u_{2}^{a}e^{-i\Psi^{(2)}_{\GR}}e^{i\beta u_{2}^b}\nn\\
&+\left[\gamma_{+}F_{+}+\gamma_{\times}F_{\times}+\gamma_{\bb}F_{\bb}+\gamma_{\LL}F_{\LL}+\gamma_{\sn}F_{\sn}+\gamma_{\se}F_{\se}\right] \eta^{1/5} u_{1}^{c}e^{-i \Psi^{(1)}_{\GR}} e^{i\delta u_{1}^d}\,.
\end{align}
\end{widetext}
Comparing this to Eq.~($\ref{ppE1}$), it is obvious that the parameters ($a,\beta, b, c, \delta, d$) are the same as introduced for a single-detector. In order to account for multiple detectors, however, we have introduced a number of parameters in place of ($\alpha, \gamma,\Phi_{c}^{(\ell)}$), one for each polarization mode: $\alpha \to (\alpha_{+},\alpha_{\times},\alpha_{\bb},\alpha_{\LL},\alpha_{\se},\alpha_{\sn})$ and $\gamma \to (\gamma_{+},\gamma_{\times},\gamma_{\bb},\gamma_{\LL},\gamma_{\se},\gamma_{\sn})$. We have here absorbed any possible inclination dependence into these parameters, as multiple detectors will essentially see the same inclination angles. Also, we have here included both $\alpha_{+}$ and $\alpha_{\times}$ for the sake of generality, but in all examples encountered usually these parameters are not independent. 

The parameters introduced are either purely real or purely imaginary, but not complex. We can then rewrite them as $\alpha_i=\bar{\alpha}_i e^{i \pi /2 \hat{\alpha}}$ and $\gamma_i=\bar{\gamma}_i e^{i \pi /2 \hat{\gamma}}$ where  $(\bar{\alpha}_i,\bar{\gamma}_i)$ are real parameters, while $(\hat{\alpha}_i,\hat{\gamma}_i)$ are  binary parameters, ie.~they are either equal to unity or they vanish. Although this introduces a new parameter $\hat{\alpha}$ or $\hat{\gamma}$ for each $\bar{\alpha}$ or $\bar{\gamma}$ parameter, the former two are not full parameters, in that they only increase the volume of the parameter space but not its dimensionality. 

As we discussed in Sec.~\ref{sec:gen-sd} and~\ref{sec:deconstruct}, however, not all of these parameters are independent. Using the arguments of Sec.~\ref{sec:deconstruct}, we can restrict the large class of templates of Eq.~\eqref{ppE1md} to the leading-order corrections induced by modified GW generation corrections without preferred frames:
\begin{widetext}
\begin{align}\label{ppE2md}
\tilde{h}_{\ppE,1}^{\MD}(f)&=\tilde{h}^{\GR}_{\MD} (1 + c \;  \beta u_2^{b+5})e^{2 i\beta u_{2}^b}+[\alpha_{\bb}F_{\bb}\sin^2{\iota}+\alpha_{\LL}F_{\LL}\sin^2{\iota}+\alpha_{\sn}F_{\sn}\sin{\iota}+\alpha_{\se}F_{\se}\sin{2\iota}]\frac{\mathcal{M}^2}{D}u_{2}^{-7/2}e^{-i\Psi^{(2)}_{\GR}}e^{2 i\beta u_{2}^b}\nn\\
&+\left[\gamma_{\bb}F_{\bb}\sin{\iota}+\gamma_{\LL}F_{\LL}\sin{\iota}+\gamma_{\sn}F_{\sn}+\gamma_{\se}F_{\se}\cos{\iota}\right] \eta^{1/5} \frac{\mathcal{M}^{2}}{D}u_{1}^{-9/2}e^{-i \Psi^{(1)}_{\GR}} e^{i \beta u_{1}^b}\,.
\end{align}
\end{widetext}
As before, there is no need to introduce $\Phi_{c}^{(\ell)}$ here. The $10$ theory parameters are then $(b,\beta,\alpha_{\bb},\alpha_{\LL},\alpha_{\se},\alpha_{\sn},\gamma_{\bb},\gamma_{\LL},\gamma_{\se},\gamma_{\sn})$. Of these, $b \in {\mathbb{Z}}$, while all others depend on the fundamental coupling constant of the theory. 

We have proposed the specific parameterization of the amplitude corrections based on the following considerations. For theories with no preferred directions and for circular orbits, the leading-order, trace-reversed metric perturbation can have terms proportional to 
\begin{enumerate}
\item For $\ell=2$ 
\\
\begin{align}
\bar{h}^{00} &\sim (\hat{N} \cdot \hat{v})^2 \,, (\hat{N} \cdot \hat{x})^2
\\
\bar{h}^{0j} &\sim (\hat{N} \cdot \hat{v}) v^{j} \,, (\hat{N} \cdot \hat{v}) x^{j} \,, (\hat{N} \cdot \hat{x}) v^{j} \,, (\hat{N} \cdot \hat{x}) x^{j}
\\
\bar{h}^{ij} &\sim (\hat{N} \cdot \hat{v})^2 \delta^{ij} \,, (\hat{N} \cdot \hat{x})^2 \delta^{ij} \,, v^i v^j \,, x^i x^j
\end{align}
\item For $\ell=1$
\\
\begin{align}
\bar{h}^{00} &\sim (\hat{N} \cdot \hat{v}) \,, (\hat{N} \cdot \hat{x})
\\
\bar{h}^{0j} &\sim v^{j} \,,  x^{j} 
\\
\bar{h}^{ij} &\sim (\hat{N} \cdot \hat{v}) \delta^{ij} \,, (\hat{N} \cdot \hat{x}) \delta^{ij} 
\end{align} 
\end{enumerate}
If we apply the projectors in Eq.~$\eqref{amplitudes}$ to the above equations to extract the polarization amplitudes and then use Eq.~$\eqref{modes}$ to calculate each mode, we find that the inclination dependence given in Eq.$\eqref{ppE2md}$ is unique for the leading-order terms. This is confirmed in the theories we discussed above, all of which lead to the same structure: the $\ell=1$ harmonic has a term proportional to $F_{A}$ and a similar term appears in the $\ell=2$ harmonic but multiplied by another factor of $\sin{\iota}$ and a different amplitude, where the latter depends on the particular theory considered. The inclination dependence will be different for terms that
enter at higher order in the orbital velocity, just as happens in GR.
We have here not allowed for ppE theory parameters multiplying $F_{+}$ and $F_{\times}$ in the $\ell=1$ harmonic because such terms could only arise if $\bar{h}_{ij} \sim \hat{x}_i \; A_j+\hat{v}_i \; B_{j}$, where $(A_i,B_i)$ are independent of the orbital phase, which would be indicative of theories with preferred directions. Such theories exist~\cite{Jacobson:2004ts,Eling:2004dk,Jacobson:2008aj}, but we leave a detailed analysis of these effects to future work. 

Finally, we can also allow for propagation modifications, in addition to the generation ones, by introducing a further phase corrections of the form
\begin{widetext}
\begin{align}\label{ppE3md}
\tilde{h}_{\ppE,2}^{\MD}(f)&=\tilde{h}^{\GR}_{\MD} (1 + c \;  \beta u_2^{b+5})e^{2 i\beta u_{2}^b}e^{i \kappa u_{2}^{k}}
\nn \\
&+
[\alpha_{\bb}F_{\bb}\sin^2{\iota}+\alpha_{\LL}F_{\LL}\sin^2{\iota}+i \alpha_{\sn}F_{\sn}\sin{\iota}+\alpha_{\se}F_{\se}\sin{2\iota}]\frac{\mathcal{M}^2}{D}u_{2}^{-7/2}e^{-i\Psi^{(2)}_{\GR}}e^{2 i\beta u_{2}^b} e^{i \kappa u_{2}^{k}}\nn\\
&+\left[\gamma_{\bb}F_{\bb}\sin{\iota}+\gamma_{\LL}F_{\LL}\sin{\iota}+i\gamma_{\sn}F_{\sn}+\gamma_{\se}F_{\se}\cos{\iota}\right] \eta^{1/5} \frac{\mathcal{M}^{2}}{D}u_{1}^{-9/2}e^{-i \Psi^{(1)}_{\GR}} e^{i \beta u_{1}^b} e^{i \kappa u_{1}^{k}}\,.
\end{align}
\end{widetext}
where $(\kappa,k)$ are propagation ppE parameters. As in Eq.~\eqref{ppE3}, using this waveform might not be ideal, as usually either $u^{b}$ or $u^{k}$ will dominate, and one might wish to neglect the subdominant contribution. 

Before proceeding, let us make a remark about the GR limit of the parameterizations discussed here. We saw earlier that both Rosen's and Lightman-Lee theory do not possess a continuous GR limit. That is, if one takes the coupling parameters of this theory to zero ${\cal{G}} \to 0$, one does not recover Einstein's theory. While our parameterization is designed to have a smooth GR limit, it does manage
to cover these exotic and observationally disfavoured alternatives.  More precisely, the parameterization discussed above poses no restriction on the magnitude of the amplitude $\beta$, and thus, if $\beta$ were large enough and $b = -5$,  the phase terms proportional to $\beta$ could combine with the GR terms to change the leading-order behavior. Such behavior can be discarded if one imposes priors on $\beta$ and $\kappa$, for example, by requiring them to satisfy their current binary pulsar constraints~\cite{Yunes:2010qb}. 

%%%%%%%%%%%%%%%%%%%%%%%%%%%%%%%%%%%
\section{Null Streams}
\label{sec:null streams}

A general method used in GW data analysis to separate signals from noise is that of null streams. This approach was first introduced by G\"ursel and Tinto~\cite{Guersel:1989th} and was later extended by Chatterji et al.~\cite{Chatterji:2006nh}. The idea behind this is to combine the data from a network of detectors to find linear combinations that contain no GW signal, only noise, in the hopes of separating false-alarms from real GW burst events. Their analysis only included the $h_{+}$ and $h_{\times}$ polarizations, since they were considering only GR events. 

We can extend this method to test Einstein's theory through the proper combination of output from several detectors. In essence, one can extend the analysis of~\cite{Guersel:1989th,Chatterji:2006nh} to allow for all possible polarizations and then construct the appropriate null streams (``null'' within GR) through appropriate projections. This method is promising to test GR, since the detection of a GW in a GR null stream would automatically signal a deviation from GR (assuming the source location is known - the more general analysis for sources with unknown sky location will be developed elsewhere). One can thus search for statistically significant deviations from noise in GR null streams, both with a template (as given in Sec.~\ref{sec:ppEGen}) and without one. Given a detection of a signal in a GR null stream, one could then reconstruct the signal through the templates in Sec.~\ref{sec:ppEGen}.

Let us assume that there exist $D \geq 6$ detectors with uncorrelated noise and that, for a given source, we know its position in the sky, as might be the case if we have an electromagnetic counterpart. Given this, one knows exactly how to time-shift the signal from detector to detector. For a detector $a$, the noise-weighted signal from a source at location $\hat{\Omega}_s$ in the sky in the frequency domain is
\be\label{data}
\tilde{d}_a=F^{+}_a\tilde{h}_{+}+F^{\times}_a\tilde{h}_{\times}+F^{\se}_a\tilde{h}_{\se}+F^{\sn}_a\tilde{h}_{\sn}+F^{\bb}_a\tilde{h}_{\bb}+F^{\LL}_a\tilde{h}_{\LL}+\tilde{n}_a\,.
\ee
Here, and in the remaining section, $\tilde{d}_a$, $F^{\cdot}_a$ and $\tilde{n}_a$ are the noise-weighted signal, the antenna patterns and the noise of the $a$th detector respectively, each defined as their standard value divided by $\sqrt{S_a(f)/2}$ where $S_a(f)$ is the power spectral density.

\begin{figure}[htb]
\centering
 \includegraphics[height=8.5cm,width=7cm]{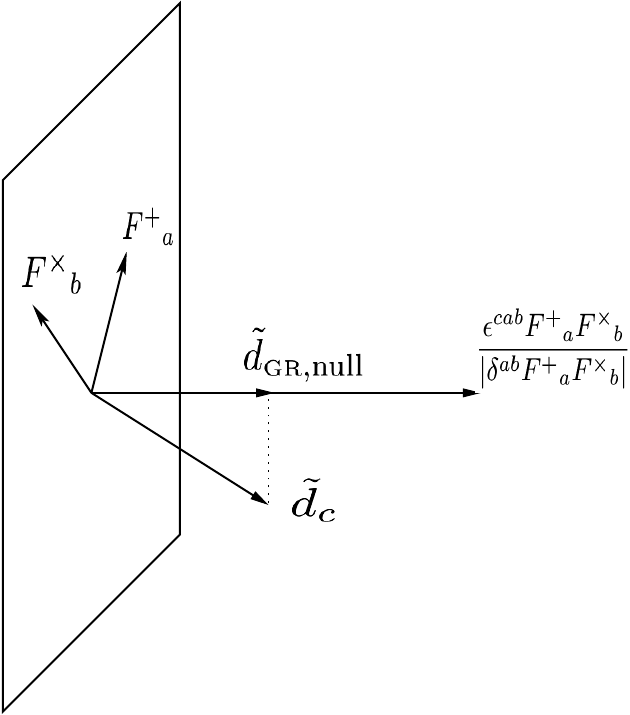}
\caption{\label{figure} Schematic diagram of the projection of the data stream $\tilde{d}_{c}$ orthogonal to the GR subspace $(F^{+}_{a},F^{\times}_{b})$ for 3 detectors to build the GR null stream $\tilde{d}_{\rm GR, null}$.}
\end{figure}

Given $D$ detectors, we can then rewrite Eq.~($\ref{data}$) as
\be
\left[ {\begin{array}{c}
 \tilde{d}_1  \\
 \tilde{d}_2  \\
  \vdots          \\
 \tilde{d}_D
\end{array} } \right]
\hspace{-0.2cm}= \hspace{-0.2cm}
\left[ {\begin{array}{cccccc}
 F^{+}_1 & F^{\times}_1 & F^{\se}_1 & F^{\sn}_1 & F^{\bb}_1 & F^{\LL}_1  \\
 F^{+}_2 & F^{\times}_2 & F^{\se}_2 & F^{\sn}_2 & F^{\bb}_2 & F^{\LL}_2  \\
  \vdots &  \vdots      &   \vdots & \vdots   &  \vdots &  \vdots     \\
 F^{+}_D & F^{\times}_D & F^{\se}_D & F^{\sn}_D & F^{\bb}_D & F^{\LL}_D
\end{array} } \right]
\left[ {\begin{array}{c}
 \tilde{h}_+  \\
 \tilde{h}_{\times}  \\
  \vdots          \\
 \tilde{h}_{\LL}
\end{array} } \right]
\hspace{-0.1cm}
+
\hspace{-0.1cm}
\left[ {\begin{array}{c}
 \tilde{n}_1  \\
 \tilde{n}_2  \\
  \vdots          \\
 \tilde{n}_D
\end{array} } \right],
\ee
or using tensor notation,
\be
\tilde{d}^a=F^a{}_j\tilde{h}^j+\tilde{n}^a\,,
\ee
where the $a$ runs over the number of detectors and the $j$ over the polarizations. The quantity $F^{a}{}_{j}$ is acting analogous to a metric in the signal manifold, and thus, the first term on the right-hand side can be interpreted as the projection of the wave vector $\tilde{h}^j$ along the directions $F^{+}{}_a, F^{\times}{}_a, F^{\se}{}_a, F^{\sn}{}_a, F^{\bb}{}_a$ and $F^{\LL}{}_a$. Therefore, one can create data sets that have no component of a certain polarization by projecting them to a direction orthogonal to the direction defined by the beam pattern functions of this polarization mode. This is illustrated in Fig~\ref{figure} for $3$ detectors.
%%%%%%%%%
%%%%%%%%%

For D detectors, the signal manifold is $D$ dimensional with $D$ basis vectors, $5$ of which can be chosen along the $F^{+}{}_a, F^{\times}{}_a, F^{\se}{}_a, F^{\sn}{}_a$ and $F^{\bb}{}_a$ directions. The reason why we cannot choose 6 linearly independent basis vectors along the $F^{.}_a$ directions is that $F^{\bb}=-F^{\LL}$, as can be clearly seen from Eq.~\ref{antpatterns} (see also~\cite{Boyle:2010gc,Boyle:2010gn}). This means that although we have a $6\times D$ matrix, this only has $5$ linearly independent columns. The remaining $D-5$ vectors would give us complete null streams, ie.~streams with truly no signals, no matter how many polarizations are present. Currently, there are only 3 detectors active or soon to be active and collecting data, and thus, we can only eliminate two polarization modes from the full signal. Let us then define the \emph{GR null stream} as the stream that has no tensor modes ($\tilde{h}_+$ and $\tilde{h}_{\times}$), namely
\be
\tilde{d}_{\GR,\rm{null}}=\frac{\epsilon^{cab}F^{+}{}_a F^{\times}{}_b}{|\delta^{ab}F^{+}{}_a F^{\times}{}_b|} \tilde{d}_{c}\,,
\ee
(this is analogous to the null stream defined in~\cite{Chatterji:2006nh}). Clearly, this will contain a mixture of the two scalar modes and two vector modes. 

Given data from $3$ detectors, one can then study whether $\tilde{d}_{\GR,\rm{null}}$ contains any statistically significant deviations from noise without ever invoking a template. However, since we have constructed a parameterization for a response function with all polarizations, one could filter $\tilde{d}_{\GR,\rm{null}}$ with the GR null scalar
\be
\tilde{h}_{\GR,\rm{null}}=\frac{\epsilon^{cab}F^{+}{}_a F^{\times}{}_b}{|\delta^{ab}F^{+}{}_a F^{\times}{}_b|} \tilde{h}_{c}\,,
\ee
fitting for the system parameters, together with the ppE theory parameters. The quantity $\tilde{h}_{c}$ here should be interpreted as, for example, $\tilde{h}_{\ppE,1}^{\MD}$ given in Eq.~\eqref{ppE1md} for the $c$th detector. 

What if we had more than $3$ detectors? Each additional detector would allow us to eliminate one of the two vector modes or both the scalar modes, beyond the two tensor ones. For example, given four detectors, $\tilde{d}_{\GR,\rm{null}}$ becomes a two vector, as there are two GR null streams, which in addition also contain no power in one of the vector modes or the scalar modes, depending on which direction we chose to project out. Therefore, given $5$ detectors one can construct $3$ enhanced GR null streams, each with power in a signal direction. With more than $5$ detectors, one can then construct complete null streams (in any theory of gravity). This topic and its implementation will be studied more carefully elsewhere~\cite{Laura:prep}.

%%%%%%%%%%%%%%%%%%%%%%%%%%%%%%%%%%%%%%%%%%%%
\section{Conclusions}
\label{sec:Conclusions}

We have studied how theories that predict the existence of all GW polarization modes affect the GW response function. First, we considered three modified theories (Brans-Dicke theory, Rosen's theory and Lightman-Lee theory) and, for each of them, we extracted the polarization modes and calculated the response function in the time and frequency domains, for a quasi-circular, non-spinning compact object inspiral source. We found that, although generically the response function contains terms proportional to the $\ell = (0,1,2)$ harmonic of the orbital phase, only the $\ell=(1,2)$ harmonics contribute significantly to the Fourier transform in the SPA. This is because the $\ell=0$ harmonic lacks a stationary point in the generalized Fourier integral, and thus, it becomes subdominant relative to the $\ell=(1,2)$ harmonics by the Riemann-Lebesgue lemma~\cite{Bender}.  

We then considered the generic structure of the $\ell$-harmonic response function in the time domain using symmetry arguments and dimensional analysis. We Fourier-transformed this generic structure in the SPA and found that the leading-order Fourier amplitude and phase scale as $\ell^{-1/2} \eta^{(2 - \ell)/5} u_{\ell}^{(2\ell - 11)/2}$ and $\ell u_{\ell}^{-5}$ respectively, where $u_{\ell} = (2 \pi {\cal{M}} f/\ell)^{1/3}$, $\eta$ is the symmetric mass, $f$ is the GW frequency and ${\cal{M}}$ is the chirp mass. We then allowed for power-law modifications to the radiation-reaction force (the energy flux) and the Hamiltonian (the binding energy). Given a modification of relative $n$th PN order to either of these quantities, we found that the Fourier amplitude and phase acquire corrections that both scale as $u_{\ell}^{2n}$ relative to the GR leading-order term. 

With this in hand, we considered the ppE framework and whether it was capable of capturing this higher-harmonic modifications to the response function. We found that, although the original ppE framework is perfectly capable of modeling the $\ell=2$ harmonic, it needs to be modified to also allow for the $\ell=1$ harmonic. We proposed two possible extensions to the single-detector ppE framework: a generic one with a total of $9$ ppE theory parameters; and a reduced one with only $4$ ppE theory parameters that uses some of the scaling relations found for a generic terms promotional to the $\ell$-harmonic of the orbital phase in the response function. 

The original ppE scheme was not applicable to multiple detector configurations, as it was based on a generic deformation of a GR {\emph{single-detector}} response. When trying to constrain the existence of non-GR polarizations, however, one needs to redo this analysis to allow for multiple detectors. We carried out such an extension proposing another two ppE, multiple-detector template families: a generic one with a total of $18$ ppE theory parameters; and a reduced one with a total of $10$ ppE parameters. The large increase in ppE parameters is necessary if one wishes to test for the existence of all $4$ additional non-GR polarizations in both the $\ell=1$ and $\ell=2$ harmonic of the Fourier transformed response function. If one restricts attention to only one additional non-GR polarization, then one requires only $9$ ppE parameters for the generic proposal and $5$ ppE parameters for the restricted proposal.

Some of the parameters we introduced here are already constrained by binary pulsar observations. For example, if there exists a scalar breathing mode at leading, Newtonian order, then it will modify the amplitude through a $-1$PN order term, which then will change the rate of change of the binding energy, and thus, of the orbital period. Such negative-PN order terms must be very small in magnitude to be compatible with binary pulsar observations~\cite{Yunes:2010qb}. Such restrictions could be accounted for in the form of a prior through a Bayesian analysis. 

We conclude with a discussion of a possible method to test for the existence of additional non-GR polarization through the combination of the output of multiple detectors. That is, we extend the concept of null streams to non-GR theories, defining a GR null stream as one that would be consistent with noise in GR, but that would contain non-GR polarizations in a modified gravity theory. Such a null stream is simply constructed by building basis vectors orthogonal to the $2$ GR polarization basis vectors in the signal manifold, and then projecting out the response with these orthogonal directions. We find that at least $3$ detectors are necessary to construct one such null stream. One requires $5$ detectors to isolate the $4$ additional non-GR polarizations into single GR null streams and $6$ detectors are required to create a null streams in all theories of gravity.
 
Although the analysis presented here is a definite step forward toward the construction of model-independent tests of GR with GWs from binary inspirals, it leaves open several directions for future research. The four ppE proposals discussed in this paper were in part inspired by our analysis of three specific models. In principle, one would like to repeat the analysis carried out here for vector-tensor theories, such as ~\cite{Will:1993ns}, stratified theories~\cite{Rosen:1971qp,Rosen:1974ua,1973PhRvD...8.3293L,Will:1993ns} and tensor-vector-scalar theories, such as TeVeS~\cite{Bekenstein:2004ne,Bekenstein:2005nv} or Einstein-Aether theory~\cite{Jacobson:2004ts,Eling:2004dk,Jacobson:2008aj}. The latter will be particularly helpful in improving the ppE scheme, as it predicts the existence of preferred directions, which should introduce $\ell=1$ harmonic modifications to the time-domain response function. Another step towards the improvement of the above proposals would be to determine a generic parametrization for the quantities that depend on the parameters of the system, like the mass, the spins etc. 

One of the main motivations for the ppE scheme is to develop a framework to test GR in the strong-field and here we have concentrated on the inspiral phase only, which might seem like a contradiction. As explained in the Introduction, however, by ``strong-field'' we here mean the region of spacetime where the gravitational field is dynamical and non-linear. The PN expansion allows one to model such non-linearities and strong dynamics perturbatively, and it is precisely this that the ppE framework modifies. Ideally, one would also want to consider extensions of the ppE framework during the plunge and merger phase. In these phases, one expects the most amount of radiation emitted through non-GR polarizations, and perhaps, the largest deviations from our GR expectations. The plunge and merger, however, are extremely difficult to model even within GR. Lacking a numerical modeling of these phases in modified gravity theories, one can only improve on inspiral ppE models at the current time. In the future, however, it would be most interesting to see how additional polarizations modify GR waveforms during merger.  

Another obvious direction for future research is the implementation of such proposals in a realistic data analysis pipeline, like that discussed in~\cite{Cornish:2011ys} or~\cite{Li:2011cg}. In particular, the amplitude of the $\ell=1$ harmonic to the Fourier transform of the response function is proportional to a ppE theory parameter, which we are assuming here to be small, relative to the GR amplitude of the $\ell=2$ mode. Since GW detectors are much more sensitive to the GW phase than its amplitude, it is not entirely clear whether the $\ell=1$ harmonic can be extracted for low signal-to-noise ratio events. Whether the inclusion of such parameter amplitude ppE parameters is of practical use can only be determined through an implementation of these proposals. 

%%%%%%%%%%%%%%%%%%%%%%%%%%%%%
\acknowledgments

We would like to thank Leo Stein and Kent Yagi for useful comments and suggestions. NY acknowledges support from NSF grant PHY-1114374 and NASA grant NNX11AI49G, under sub-award 00001944. NC acknowledges
support from NSF award 0855407 and NASA grant NNX10AH15G.

%%%%%%%%%%%%%%%%%%%%%%%%%%%%%
%%%%%%%%%%%%%%%%%%%%%%%%%%%%%
\bibliography{review}

%%%%%%%%%%%%%%%%%%%%%%%%%%%%%

\section{Erratum: Model-Independent Test of General Relativity:  An Extended post-Einsteinian Framework with Complete Polarization Content [Phys.~Rev.~D 86, 022004 (2012)]}
%OTHER OPTIONS:
%Improved post-Einsteinian framework:
%Complete polarization content, many detectors case

%%%%%%%%%%%%%%%%%%%%%%%%%%%%%%%%%%%%%%%%%%%%%%%
\begin{center}

In this erratum, we correct a mistake in the deconstruction of the inspiral waveform generation in the parameterized post-Einsteinian framework. These corrections do not affect the conclusions of the paper.

\end{center}

%%%%%%%%%%%%%%%%%%%%%%%%%%%%%%%%%%%%%%%%%%%%%%%
%\pacs{04.80.Cc,04.80.Nn,04.30.-w,04.50.Kd}

%\maketitle

%%%%%%%%%%%%%%%%%%%%%%%%%%%%%%%%%%%%%%%%%%%%%%%
\section{Deconstruction of Inspiral Waveform Generation}
\label{sec:deconstruct}

In~\cite{Chatziioannou:2012rf}, we described how to construct a gravitational waveform due to the inspiral of two compact objects in General Relativity and in the parameterized post-Einsteinian (ppE) framework. We recently discovered a few mistakes in Sec.~IV of this paper that we correct here. None of these corrections affect the conclusions of our work, but they do affect some of the intermediate expressions, as we describe below. 

We begin through the ppE correction to the reduced effective potential:
\begin{align}
V_{\rm eff} &= \left(- \frac{m}{r} + \frac{L_{z}^2}{2 r^2} \right) \left[1 + A \left(\frac{M}{r}\right)^p\right]\,,
\end{align}
where $m$ is the total mass of the system, $r$ is the orbital separation, $L_{z}$ is the z-component of the angular momentum, and $A$ and $p$ are ppE amplitude and exponent modifications. Taking the radial derivative of this effective potential and setting it to zero, we find the modified Kepler's law, as given in Eq.~$(91)$ of~\cite{Chatziioannou:2012rf}. Since the effective potential is, in this case, equal to the binding energy (since the radial kinetic energy term is zero for circular orbits), we find 
\begin{align}
E_{b} &= - \frac{1}{2}  \eta^{-2/5} \left(2 \pi {\cal{M}} F\right)^{2/3} 
\nn \\
&\times  \left[1 - \frac{1}{3} A \left(2 p - 3\right) \eta^{-2 p/5} \left(2 \pi {\cal{M}} F\right)^{2p/3}\right]\,,
\end{align}
where ${\cal{M}}$ is the chirp mass, $F$ is the orbital frequency and $\eta$ is the symmetric mass ratio. This expression corrects Eq.~$(94)$ in~\cite{Chatziioannou:2012rf}.

From the corrected expression for the binding energy presented above, we can now compute the rate of change of the orbital frequency with time: 
\begin{align}
\frac{dF}{dt} &= \frac{48}{5\pi{\cal{M}}^{2}}  \left(2 \pi {\cal{M}} F\right)^{11/3} \left[1 
\right.
\nn \\
&+\left. 
B \eta^{-2q/5} \left(2 \pi {\cal{M}} F\right)^{2q/3} 
\right.
\nn \\
&+\left. 
\frac{1}{3} A  \left(2 p^{2} -2 p - 3\right) \eta^{-2p/5} \left(2 \pi {\cal{M}} F\right)^{2p/3}\right]\,,
\end{align}
which corrects Eq.~(98) in~\cite{Chatziioannou:2012rf}. Notice that there was an inconsequential typo in the in-line equation between Eqs.~(51) and (52)~\cite{Chatziioannou:2012rf}, where the binding energy was missing a factor of $1/2$. Notice also that in deriving $dF/dt$ above, we used the rate of change of the binding energy (i.e.~the energy flux), as written in Eq.~(96) of~\cite{Chatziioannou:2012rf}. When writing this flux as a function of the orbital frequency, we implicitly assumed that $\dot{E}_{\rm GR}$ is proportional to $v^{2} (m/r)^{4}$. If one assumes a different form for the GR expression, such as one that is proportional to $r^{4} \omega^{6}$, one will find slightly different expressions for the waveform due to the ppE modifications to Kepler's third law.

Propagating these corrections further, we encounter the Fourier transform of the ($\ell$-harmonic of the) gravitational waveform. The latter is define in Eqs. (17) and (18) of~\cite{Chatziioannou:2012rf}, which have an inconsequential typo since both expressions should depend on the second derivative of the quadrupole moment $\ddot{Q}_{ij}$ instead of $Q_{\ij}$. The corrected Fourier transform of the waveform is
\begin{align}
\tilde{h}^{(\ell)} &= \sqrt{\frac{5 \pi}{48 \ell}} Q \frac{{\cal{M}}^{2}}{D} \eta^{(2 - \ell)/5} u_{\ell}^{(2 \ell -11)/2} e^{-i \Psi^{(\ell)}}
\nn \\
&\times
\left[1 - \frac{1}{2} B \eta^{-2q/5} u_{\ell}^{2q} 
\right.
\nn \\
&+ \left. 
\frac{1}{6} A  \left(\ell p + 3 + 2 p - 2 p^{2}\right) \; \eta^{-2p/5} \; u_{\ell}^{2 p} \right]\,,
\label{hofell}
\end{align}
correcting Eq.~(99) in~\cite{Chatziioannou:2012rf}.  Here $Q$ is a function of some sky angles, $u_{\ell}$ is a reduced frequency, $D$ is the distance to the source, and $B$ and $q$ are ppE amplitude and exponent corrections to the energy flux (all of which are defined in more detail in~\cite{Chatziioannou:2012rf}).  Similarly, the corrected phase of the Fourier transform is
\begin{align}
\delta\Psi^{(\ell)} &= \frac{5}{64} A \; \frac{\ell \left(2 p^{2} - 2 p - 3\right)}{(4-p)(5-2p)} \eta^{-2 p/5} u_{\ell}^{2p-5} 
\nn \\
&+
\frac{15}{64} B \frac{\ell}{(4-q)(5-2q)} \eta^{-2q/5} u_{\ell}^{2q-5}\,,
\label{deltapsi}
\end{align}
which corrects Eq.~(100) in~\cite{Chatziioannou:2012rf}. Finally, the correction to the Fourier phase at the $p=4$ pole remains equal to Eq.~(100) in~\cite{Chatziioannou:2012rf} but with the prefactor $280/9$ instead of $880/9$, while that at $p=5/2$ remains equal to Eq.~(103) in~\cite{Chatziioannou:2012rf} but with the prefactor $20/3$ instead of $30$. 

These modifications to the inspiral gravitational waveform then lead to modifications to the mapping between ppE coefficients and ppE modifications to the binding energy. In particular, Eqs.~(106) and (107) in~\cite{Chatziioannou:2012rf} are now corrected to
\begin{align}
\alpha &= \left[\frac{1}{6} A \left( 3 + 4 k - 2 k^{2}\right) - \frac{1}{2} B\right] \eta^{-2k/5}\,,
\\
\beta &= -\frac{5}{64} \frac{1}{(4-k)(5-2k)} \left[A \left(2k^{2} - 2 k - 3\right) + 3 B \right]\eta^{-2k/5}\,,
\end{align}
and Eqs.~(109) and (110) in~\cite{Chatziioannou:2012rf} are now corrected to
\begin{align}
\alpha &= \frac{1}{6} A \left(3 + 4p  - 2 p^{2}\right) \eta^{-2p/5}\,,
\\
\beta &= -\frac{5}{64} \frac{1}{(4-p)(5-2p)} A \left(2p^{2} - 2 p - 3\right) \eta^{-2p/5}\,,
\end{align}
Similarly, Eq.~(118)  in~\cite{Chatziioannou:2012rf} becomes
\be
\label{c-cons}
c_{\rm{cons}} = -\frac{8}{15} \frac{b(3 - b) (b^{2} + 6 b -1)}{b^{2} + 8 b + 9}\,,
\ee
and Eq.~(120) in~\cite{Chatziioannou:2012rf} becomes
\be
c_{\rm{both}} = -\frac{16}{15} \frac{b(3 -b) (b^{2}  + 7 b + 4)}{b^{2} + 8 b + 9}\,.
\label{c-both}
\ee
%

%
%\footnotesize{.} 
%
%\includepdf[
%  pages={1},
%  scale=1,clip,trim=0cm 0cm 0cm 0cm,
% ]{paper.pdf}
% 
%\newpage
% \includepdf[
%  pages={2},
%  scale=1,clip,trim=0cm 0cm 0cm 0cm,
% ]{paper.pdf}

\end{document}